\RequirePackage[2020-02-02]{latexrelease}
\documentclass[11pt,a4paper,english,superscriptaddress,aps]{revtex4}

\usepackage{color}
\usepackage{amsmath}
\usepackage{amssymb,amsfonts,latexsym,cancel} 
\usepackage{graphicx}
\usepackage{hhline}
\usepackage{mwe}
\usepackage{amsbsy}
\usepackage{textcomp}
\usepackage{commath}
\usepackage{subfig}
\usepackage{float} 
\usepackage{mathrsfs}
\usepackage{bm}
\usepackage{array} 
\usepackage{subfig}
\usepackage[margin=0.9in]{geometry}

\DeclareMathOperator{\csch}{csch}
\DeclareMathOperator{\sech}{sech}
\DeclareMathOperator{\sinc}{sinc}
\DeclareMathOperator{\arctanh}{arctanh}

\usepackage{babel}
\newcommand{\bea}{\begin{eqnarray}}
\newcommand{\eea}{\end{eqnarray}}

\newcommand{\be}{\begin{equation}}
\newcommand{\ee}{\end{equation}}

\providecommand{\abs}[1]{\lvert#1\rvert}

\numberwithin{equation}{section}

\makeatletter

\begin{document}

\title{Inclusion of radiation in the CCM approach of the $\phi^4$ model}
\date{\today}

\author{S. Navarro-Obreg\'on}
\email{sergio.navarro.obregon@uva.es}
\author{L.M. Nieto}
\email{luismiguel.nieto.calzada@uva.es}

\affiliation{Departamento de F\'{\i}sica Te\'orica, At\'omica y \'Optica,
Universidad de Valladolid, 47011 Valladolid, Spain}

\author{J.M. Queiruga}
\email{xose.queiruga@usal.es}
\affiliation{Department of Applied Mathematics, University of Salamanca,
Casas del Parque 2, 37008 - Salamanca, Spain}
\affiliation{Institute of Fundamental Physics and Mathematics, University of Salamanca, Plaza de la Merced 1, 37008 - Salamanca, Spain}

\begin{abstract}
We present an effective Lagrangian for the $\phi^4$ model that includes radiation modes as collective coordinates. The coupling between these modes to the discrete part of the spectrum, i.e., the zero mode and the shape mode, gives rise to different phenomena which can be understood in a simple way in our approach. In particular, the energy transfer between radiation, translation and shape modes is carefully investigated in the single-kink sector. Finally, we also discuss the inclusion of radiation modes in the study of oscillons. This leads to relevant phenomena such as the oscillon decay and the kink-antikink creation.
\end{abstract}

\maketitle


\section{Introduction}

Topological solitons are non-linear field theory solutions that appear in many branches of physics, from condensed matter to cosmology \cite{Rajaraman, Manton, Shellard, Shnir}, and that have gained interest over the last decades. The stability of these objects is guaranteed by their topological charge, which is a topological invariant conserved during time evolution. They have particle-like behaviour, so they can interact with each other, with external fields or with radiation, as well as being annihilated and even created in pairs. Among all of them, the $\phi^4$ model is particularly interesting: it can be formulated in $1+1$ dimensions, which makes it simpler from a computational point of view. In addition, the static solitons (called kinks) as well as the spectrum of perturbations can be determined analytically. The study of low-energy perturbations around the kinks shows that, in the linearised theory, there exists a discrete spectrum formed by two states localised around the core of the topological soliton - the zero mode and the shape mode - as well as a continuum of states representing radiation moving away from the kink.

However, the analysis of the dynamics in the full non-linear theory is extremely involved, which often means that the equations have to be solved numerically. This complexity comes, in part, from the multiple interaction channels, namely: attractive or repulsive static forces, excitation of the internal degrees of freedom, and interaction with radiation. One method to reduce the complexity of the topological soliton dynamics is through the Collective Coordinate Method (CCM). Within this approach, the field theory Lagrangian is reduced to a mechanical one with a finite number of degrees of freedom. In Bogomolnyi-Prasad-Sommerfield (BPS) models, this approach gives rise to the so-called canonical moduli space. Here, the relevant degrees of freedom are the positions of the solitons, and the dynamics can be described effectively as a geodesic motion in the manifold given by the BPS solitons. When we deal with non-BPS sectors this study requires the introduction of a potential, which accounts for the static interactions between solitons, but the formalism is basically the same as in the BPS sector. 

This effective point of view can be improved by introducing new coordinates which take into account internal degrees of freedom. But, even in the apparently simple case of $\phi^4$, there is a complicated pattern of final states in scattering processes related to the non-integrability of the model, which cannot be explained satisfactorily by a simple choice of translational and internal oscillatory degrees of freedom \cite{Sugiyama, Campbell}. Recently,  an important improvement has been made by means of the introduction of the relativistic moduli space \cite{Relativistic, Relativistic1}. This approach, unlike the standard CCM, can also accommodate some relativistic degrees of freedom. As shown in \cite{Relativistic}, this quantitatively improves the agreement between the effective model and the field theory. However, neither of these approaches considers dissipative degrees of freedom as generalised coordinates, i.e., they cannot describe radiation. The effect of radiation in soliton dynamics can be very relevant in certain violent processes, such as the kink-antikink annihilation, but it is also determinant in long-time dynamics and may contribute to the fractal structure of the kink scattering. One of the purposes of this work is to introduce the radiation modes as generalised coordinates and study their role in certain dynamical processes.

Apart from the topologically non-trivial solutions of the $\phi^4$ model, there are other time dependent soliton-like structures that deserve special attention, the oscillons. They are topologically trivial solutions (they are in the topological sector of the vacuum) that are long-lived, and in contrast to other time-dependent solitons such as Q-balls, they are not associated to any conserved charge. They are ubiquitous in a wide range of models from one to three dimensions \cite{Segur, Fodor1, Fodor3, Hindmarsh2, Hindmarsh3, Manton3}, and they have found applications in many scenarios in theoretical physics, from dark matter \cite{Olle,Kawasaki, Arvanitaki} to cosmology \cite{Hindmarsh5, Gorghetto,Blanco}. Some of their characteristics such as profiles and life-time have been studied, mostly numerically, in the literature. Much less is known about their internal structure (see \cite{dissel} for a recent publication). We will introduce radiation degrees of freedom in an effective model for the $\phi^4$ oscillon. They are able to provide a decay channel for oscillons below the critical amplitude. In addition, the scattering modes, or more precisely, an effective version of them, are able to describe qualitatively some features of the internal modes hosted by the oscillon, including the decay into kink-antikink pairs. 

This paper is organised as follows. In Sec. \ref{phimodel} we briefly review the $\phi^4$ model and its spectrum of perturbations. In Sec. \ref{wobbling} we introduce the radiation modes as collective coordinates and analyse the radiation emitted by a wobbling kink. After that, in Sec. \ref{Oscil_Radi} we explore some analytic solutions to the lowest order in perturbation theory involving radiation. In Sec. \ref{Trans_Oscil_Radi} we introduce the zero mode as a collective coordinate and study its interaction with the rest of the modes. In Sec. \ref{oscillon} we extend our approach to describe oscillons. Finally, Sec. \ref{conclusions} contains our conclusions and further comments. We also add two appendices with some computational details.


\section{The model and the linearised spectrum}
\label{phimodel}

The field theory model that we will discuss is the so-called $\phi ^4$ model, which is described by the following $(1 + 1)$ dimensional Lagrangian for a real scalar field $\phi(x,t)$
\begin{equation}\label{action}
\mathcal{L} =  \dfrac{1}{2}\partial_{\mu}\phi\partial^{\mu}\phi - \dfrac{1}{2}(\phi^2 - 1)^2\,.
\end{equation} 
 The field equation for this model reads 
\begin{equation}\label{eq_motion:phi4}
\square\phi + 2\phi(\phi^2 - 1) = 0\,.
\end{equation}
It can be shown by using a Bogomolnyi rearrangement that the static field configurations also satisfy the following first order BPS equations 
\be
\phi'(x)\pm (\phi(x)^2-1)=0\,.
\ee
In addition to vacuum solutions ($\phi(x)=\pm 1$), there are non-trivial solutions interpolating the vacua which can be computed analytically 
\begin{equation}
\phi_{K (\bar{K})}(x) = \pm \tanh(x-x_0)\,.
\end{equation}
\noindent
The solution with positive sign is called kink ($K$), and the one with negative sign is called antikink ($\bar{K}$). They depend on a free parameter, $x_0$, which is interpreted as the position of the kink (antikink). There is a topological charge associated to these solutions, $Q=\pm1$, which is conserved during time evolution. The perturbations around the kink (antikink) do not depend on the position or the topological charge of the solution, therefore we will consider a kink centred at the origin perturbed as follows
\begin{equation}\label{eq_perturbation}
\phi(x,t) = \phi_{K}(x) + \eta(x,t),
\end{equation}
with $\eta(x,t) = \eta(x)e^{i \omega t}$. Substituting (\ref{eq_perturbation}) into (\ref{eq_motion:phi4}), at linear order, the field equation looks like
\begin{equation}\label{eq_linearised}
-\eta''(x) + \left( 6\phi_{K}(x)^2 - 2 \right)\eta(x) = \omega^2\eta(x). 
\end{equation}

The equation (\ref{eq_linearised}) can be considered a Sturm–Liouville differential equation, where we can identify $U(x) = 6\phi_{K}(x)^2 - 2$ with a Pöschl-Teller potential. The spectrum of eigenstates and eigenvalues associated to this Schrödinger-like equation is
\begin{eqnarray}\label{linear_perturbations}
\eta_{0}(x) &=& \frac{\sqrt{3}}{2}\sech^2 x, \qquad \omega_{0} = 0, \\
\eta_{s}(x) &=& \sqrt{\dfrac{3}{2}}\sinh x\sech^2 x, \qquad \omega_{s} = \sqrt{3}, \\
\eta_{q}(x) &=&  \dfrac{ 3 \tanh^2 x -q^2 - 1 - 3iq\tanh x}{\sqrt{(q^2+1)(q^2+4)}} e^{iqx}, \qquad \omega_{q} = \sqrt{q^2 + 4},
\end{eqnarray}
with $q \in \mathbb{R}$. Altogether, the following relations are satisfied:
\begin{eqnarray}
\left\langle \eta_{0}(x),\eta_{s}(x) \right\rangle &=&\left\langle \eta_{0}(x),\eta_{q}(x) \right\rangle =
\left\langle \eta_{s}(x),\eta_{q}(x) \right\rangle = 0,\label{ortho_shape_etaq}\\
\left\langle \eta_{q}(x),\eta_{q'}(x) \right\rangle &=&2 \pi  \, \delta (q-q')\label{ortho_etaq}. 
\end{eqnarray} 
We have chosen the normalization of the scattering modes such that asymptotically they are plane waves of amplitude one. Moreover, the general theory of the Sturm–Liouville systems ensures that the eigenstates of (\ref{eq_linearised}) form a basis, so $\mathcal{B} = \left\lbrace \eta_{0}(x), \eta_{s}(x), \eta_{q}(x) \right\rbrace $ may be used to build a general configuration belonging to the linearised field configuration space. As a consequence, a general field configuration close to the kink solution can be expanded as follows
\begin{equation}\label{general_ansatz}
\phi (x,t) = \phi_{K}(x) + c_{0}(t)\eta_{0}(x) + c_{s}(t)\eta_{s}(x) + \int_{\mathbb{R}} dq\, c_{q}(t)\eta_{q}(x). 
\end{equation}
This natural assumption contains all possible degrees of freedom of the kink: the zero mode $\eta_{0}(x)$ is responsible for the infinitesimal rigid translation of the kink, the shape mode $\eta_{s}(x)$ is responsible for the modification of the width of the kink, and the radiation modes (or scattering states) $\eta_{q}(x)$ are related to the continuum of perturbative fluctuations around the vacuum, that propagate freely to infinity. Such a general ansatz will be used as the basis for the field configurations that we will study throughout this work.


\section{Leading radiation from the wobbling kink}
\label{wobbling}

In this section we study in detail the radiation emitted by a kink whose shape mode is excited with a small amplitude. A similar analysis was performed in \cite{Manton2},  albeit with a different approach, so we will use those results as a first check of the validity of our guess for the dissipative modes. In order to do that, let us assume that the kink is at rest at the origin so that we can disregard the translational degree of freedom, $\eta_0(x)$. Hence, we consider the following simplified ansatz
\begin{equation}\label{ansatz:oscill}
\phi (x,t) = \phi_{K}(x) + c_{s}(t)\eta_{s}(x) + \int_{\mathbb{R}} dq\, c_{q}(t)\eta_{q}(x)\,. 
\end{equation}
Since the shape mode solution is exact at linear order in perturbation theory, we will assume that $c_{q}(t) \sim \mathcal{O}\left(c_{s}^{2}(t)\right)$. Substituting (\ref{ansatz:oscill}) in (\ref{eq_motion:phi4}) we obtain, at linear order in $c_{s}(t)$,
\begin{equation}
\eta_{s}(x)\bigg(\ddot{c}_{s}(t) + \omega_{s}^{2}c_{s}(t)\bigg) = 0\,.
\end{equation}
Consequently, the shape mode oscillates with frequency $\omega_{s}$, i.e.,
\begin{equation}\label{harmon_oscill}
\ddot{c}_{s}(t) + \omega_{s}^{2}c_{s}(t) = 0 \Rightarrow c_{s}(t) = A_{0}\cos (\omega_{s}t)\,.
\end{equation}
Since this solution solves exactly the equation at this order, we may conclude that there is no source for radiation at linear order in the shape mode amplitude. At second order in $c_{s}(t)$ we get
\begin{equation}\label{second_order}
\eta_{s}(x)\bigg( \ddot{c}_{s}(t) + \omega_{s}^{2}c_{s}(t) \bigg) + \int_{\mathbb{R}}dq\, \eta_{q}(x)\bigg( \ddot{c}_{q}(t) + \omega_{q}^{2}c_{q}(t)\bigg) + 6c_{s}^{2}(t)\phi_{K}(x)\eta_{s}^{2}(x) = 0\,.
\end{equation}
Projecting onto $\eta_{s}(x)$ and assuming the orthogonality relations (\ref{ortho_shape_etaq}), equation (\ref{second_order}) reduces to
\begin{equation}\label{anharmon_oscill}
\ddot{c}_{s}(t) + \omega_{s}^{2}c_{s}(t) + \frac{9\pi \sqrt{6}}{32}\, c_{s}^{2}(t) = 0\,.
\end{equation}

This is the equation of an anharmonic oscillator corrected by a quadratic term. If we now project onto $\eta^{\ast}_{q'}(x)$ and assume again the orthogonality relations (\ref{ortho_shape_etaq})-(\ref{ortho_etaq}), the equation (\ref{second_order}) reads as
\begin{equation}\label{eq_etas:1}
\ddot{c}_{q}(t) + \omega_{q}^{2}c_{q}(t)  + \dfrac{3}{\pi}c_{s}^{2}(t)\int_{\mathbb{R}}dx\, \eta^{\ast}_{q}(x) \phi_{K}(x)\eta_{s}^{2}(x) = 0\,,
\end{equation}
where the last term can be interpreted as the overlap between the scattering state of frequency $\omega_{q}$ with the combination $\phi_{k}(x)\eta_{s}^2$. Such a term can be computed exactly:
\begin{equation}
\mathcal{F}(q) = \int_{\mathbb{R}}dx\, \eta^{\ast}_{q}(x) \phi_{K}(x)\eta_{s}^{2}(x) = - \dfrac{i \pi}{32} \sqrt{\dfrac{q^2 + 4}{q^2 + 1}}  \dfrac{q^2 (q^2 - 2)}{\sinh \left( \pi q/2 \right)}\,.  
\end{equation}
Therefore, (\ref{eq_etas:1}) looks like
\begin{equation}\label{eq_etas:2}
\ddot{c}_{q}(t) + \omega_{q}^{2}c_{q}(t) + \dfrac{3}{\pi}c_s^2(t)\mathcal{F}(q) = 0\,.
\end{equation}
Let us assume that the amplitude of the shape mode $c_{s}(t)$ is given by its linear approximation, i.e., $c_{s}(t) = A_{0}\cos(\omega_{s} t)$. We are considering that the shape mode is the only source of radiation, so in the absence of the shape mode there is no radiation. We will take this into account imposing the initial conditions $c_{q}(0) = 0$ and $\dot{c}_{q}(0) = 0$. With this choice, the general solution of (\ref{eq_etas:2}) takes the form
\begin{equation}\label{eq_cq}
c_{q}(t) = - \dfrac{3}{2\pi}\dfrac{(4\omega_{s}^{2} - \omega_{q}^{2}) - \omega_{q}^{2}\cos(2 \omega_{s} t) - (4\omega_{s}^2 - 2\omega_{q}^{2})\cos(\omega_{q} t)}{\omega_q^2 (4\omega_s^2 - \omega_q^2)}\mathcal{F}(q)\,.
\end{equation}
This expression provides time-dependent amplitudes to radiation modes. As a consequence, the radiation emitted by an oscillating kink is 
\begin{equation}\label{radiation}
R(x,t) = \int_{\mathbb{R}}\, dq\, c_{q}(t)\eta_{q}(x),
\end{equation}
with $c_q(t)$ given by (\ref{eq_cq}). This is the exact form of the radiation at leading order for a static wobbling kink valid at all distances and times. Notice that the structure of $\mathcal{F}(q)$ indicates that some frequencies are suppressed in the radiation. It has a maximum at $q\approx 2\sqrt{2}$, which is consistent with the fact that the shape mode is the quadratic source for radiation. Moreover, it can be shown analytically that at large distances all frequencies but $q=2\sqrt{2}$ are suppressed. A non-trivial calculation (see  Appendix \ref{AppendixA} for details) leads to the following expression for radiation at infinity   
\be\label{rad_inf}
R_\infty(x,t)= \dfrac{3\,\pi A_0^2}{2 \sinh(\sqrt{2}\pi)}\sqrt{\frac{3}{8}}\cos\big(2\sqrt{3}t - 2\sqrt{2}x - \delta\big).
\ee
This expression agrees with the one obtained in \cite{Manton2}. Following \cite{Manton2}, the decay of the amplitude of the shape mode into radiation can be determined directly from (\ref{rad_inf}). Taking into account that the average energy flux carried by the wave (\ref{rad_inf}) has to be equal to the rate of change of the energy of the excited kink (for details see \cite{Manton2}),  we arrive at
\bea\label{eq:manton_law}
A(t) = \dfrac{1}{\sqrt{ A_{0}^{-2} + 0.03\, t}}\,,
\eea
where $A_0$ is the initial amplitude of the shape mode. Notice that in this approximation there is no back-reaction of the radiation into the shape mode. Therefore, strictly speaking, the shape mode oscillates harmonically with frequency $\omega_s$ and constant amplitude. In addition, this approximation suggests that it is possible to excite resonantly the shape mode with radiation of the appropriate frequency. In order to obtain these results, we have to consider the shape mode amplitude as a free collective coordinate interacting with the radiation coordinates. We will study these issues in detail in the next section.


\section{Interaction of radiation and shape mode}\label{Oscil_Radi}

In this section we will derive, within the collective coordinate approach, the effective Lagrangian involving the shape mode coupled to radiation, i.e., $c_s(t)$ and $c_q(t)$ will be consider genuine collective coordinates. In order to achieve that, let us suppose a perturbation of the kink solution of the form (\ref{ansatz:oscill}), where $c_{s}(t) \sim \mathcal{O}(A_{0})$ and $c_{q}(t) \sim \mathcal{O}(A_{0}^{2})$ with $\abs{A_{0}} \ll 1$. Substituting (\ref{ansatz:oscill}) into (\ref{action}) we obtain
\begin{eqnarray}\label{expansion_action:oscill}
\mathcal{L}_{s,q} &=& - \dfrac{4}{3} + \dfrac{1}{2} \left( \dot{c}_{s}^{2}(t) - \omega_{s}^{2}c_{s}^{2}(t)\right) +  \pi \int_{\mathbb{R}} dq\, \left( \dot{c}_{q}(t)\dot{c}_{-q}(t) - \omega_{q}^{2}c_{q}(t)c_{-q}(t) \right) \nonumber\\
& & -  \int_{\mathbb{R}} dx\,\bigg( 2\phi_{K}(x)c_{s}^{3}(t)\eta_{s}^{3}(x) +  6\phi_{K}(x)c_{s}^{2}(t)\eta_{s}^{2}(x)\int_{\mathbb{R}} dq\,c_{q}(t) \eta_{q}(x) \bigg)\nonumber \\
& & - 	\int_{\mathbb{R}} dx\bigg(6 \phi_K(x)c_s(t)\eta_s(x)\int_{\mathbb{R}^2}dq dq' c_q(t)c_{q'}(t)\eta_q(x)\eta_{q'}(x)\bigg),
\end{eqnarray}
where the normalization of the shape mode and the delta-normalization relation (\ref{ortho_etaq}) were taken into account. Integrating in the $x-$variable we get
\begin{eqnarray}
&&\int_{\mathbb{R}} dx \, \phi_{K}(x)\,\eta_{s}(x)^{3} =\dfrac{3}{32}\sqrt{\dfrac{3}{2}} \pi\,,\\
&&\int_{\mathbb{R}} dx \, \phi_{K}(x)\,\eta_{s}^{2}(x)\,\eta_{q}(x) = \dfrac{i \pi}{32} \sqrt{\dfrac{q^2 + 4}{q^2 + 1}}  \dfrac{q^2 (q^2 - 2)}{\sinh \left( \pi q/2 \right)}\,,\\
&&\int_{\mathbb{R}}dx\, \phi_K(x)\eta_s(x)\eta_q(x)\eta_{q'}(x) =\nonumber\\
&&\frac{\pi}{16}\sqrt{\frac{3}{2}}\dfrac{17+17q^2+17 q'^2+10 q^2q'^2-q^4-q'^4+q^2q'^4+q^4q'^2-q^6-q'^6}{\sqrt{(q^2+1)(q^2+4)}\sqrt{(q'^2+1)(q'^2+4)}\cosh(\frac{\pi}{2}(q+q'))}\,. \label{cs_cq_cq}
\end{eqnarray}
One finally obtains the following expression
\begin{eqnarray}\label{effective_lagran:oscill}
\mathcal{L}_{s,q} &=& \dfrac{1}{2}\bigg(\dot{c}_{s}^{2}(t) - \omega_{s}^{2}c_{s}^{2}(t)\bigg) + \pi \int_{\mathbb{R}} dq\, \left( \dot{c}_{q}(t)\dot{c}_{-q}(t) - \omega_{q}^{2}c_{q}(t)c_{-q}(t) \right) - \dfrac{3 \pi}{16}\sqrt{\dfrac{3}{2}}c_{s}^{3}(t) \nonumber\\ 
& &  - \dfrac{3 \pi i}{16} \int_{\mathbb{R}}\, dq\, \sqrt{\dfrac{q^2 + 4}{q^2 + 1}}\dfrac{q^2 (q^2 - 2)}{\sinh (\pi q/2)} c_{s}^{2}(t)c_{q}(t) + c_s(t)\int_{\mathbb{R}^2}dq dq'\, f_{sq}(q,q') c_q(t)c_{q'}(t)\,,
\end{eqnarray}
where we have removed from (\ref{expansion_action:oscill}) a constant term (kink rest energy with opposite sign) since it does not contribute to the field equations, and $f_{sq}(q,q')$ can be read from (\ref{cs_cq_cq}). The Lagrangian (\ref{effective_lagran:oscill}) describes a system of harmonic oscillators coupled by the last two terms. The equations of motion governing the evolution of $c_{q}(t)$ and $c_{s}(t)$ are yielded by
\bea\label{effec_eq:cq}
\ddot{c}_{-q}(t) &+& \omega_{q}^{2}\,c_{-q}(t) +  \dfrac{3 i}{32}\sqrt{\dfrac{q^2 + 4}{q^2 + 1}}\dfrac{q^{2}(q^2-2)}{\sinh\left( \pi q / 2 \right)}c_{s}^{2}(t) - \dfrac{1}{\pi} c_s(t)\int dq' f_{sq}(q,q')c_{q'}(t) = 0,\\ \label{effec_eq:cs}
\ddot{c}_{s}(t) &+& \omega_{s}^{2}c_{s}(t) + \dfrac{9\pi}{16}\sqrt{\dfrac{3}{2}}c_{s}^{2}(t) + \dfrac{3\pi}{8}i \int_{\mathbb{R}} dq\, \sqrt{\dfrac{q^2 + 4}{q^2 + 1}}\dfrac{q^2 (q^2 - 2)}{\sinh( \pi q/2)} c_{s}(t)c_{q}(t) \nonumber\\
&-& \int_{\mathbb{R}^2}dq dq'\, f_{sq}(q,q') c_q(t)c_{q'}(t) = 0.
\eea

This coupled system has a straightforward interpretation: Eq. (\ref{effec_eq:cq}) is a forced harmonic oscillator of fundamental frequency $\omega_q$ with external force proportional to $c_s^2(t)$ (notice that this equation was derived in the previous section, but now we allow for back-reaction of radiation in the shape mode). Regarding Eq. (\ref{effec_eq:cs}), its structure is more involved. It describes an anharmonic oscillator of fundamental frequency $\omega_s$ coupled linearly to the $c_q(t)$. Note that, for sufficiently small amplitudes of $c_s(t)$, this expression reduces to the well-known Hill's equation provided that $c_q(t)$ is periodic. 

In order to solve the system (\ref{effec_eq:cq})-(\ref{effec_eq:cs}) numerically, we have to choose a discretization in $q$, i.e., we have to select $N$ scattering modes labelled by $q_i$ and solve the coupled system of $N+1$ ordinary differential equations. Note that the discretization of the integrals in $q$ gives rise a Hamiltonian system that conserves energy. Then, the discretization fixes a time cut-off of order $t_c=1/\Delta q$ (where $\Delta q=q_i-q_{i-1}$), which implies that for times bigger than $t_c$ our computations are no longer trustable. We will see in Sec. \ref{oscillon} that there are other effective ways to mimic the effect of dissipative degrees of freedom.

\begin{figure}[H]
\begin{center}
\includegraphics[scale=0.9]{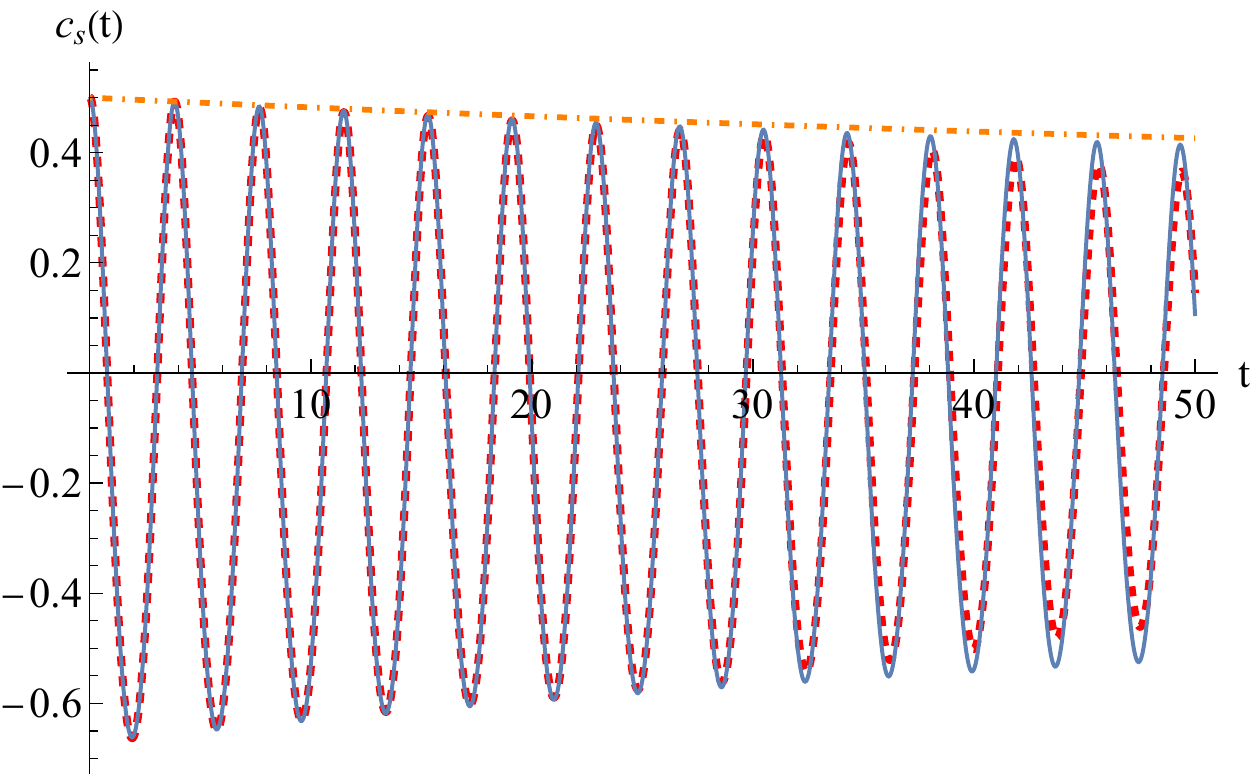}    
\caption{\small Decay of the shape mode in field theory (solid line) and in the effective model (\ref{effective_lagran:oscill}) (dashed line) compared to the analytical decay law given by (\ref{eq:manton_law}) (dotted-dashed line). For the computation we have chosen $n = 20$ equidistant scattering modes in the interval $q \in [-3,3]$.}
\label{fig:manton}
\end{center}
\end{figure}

In our first numerical experiment we will show that the Lagrangian (\ref{effective_lagran:oscill}) describes accurately the decay of the shape mode for $t \lesssim t_c$. We will choose the following initial conditions (IC)
\be
c_s(0)=A_0\,,\quad c_s'(0)=0\,,\quad c_q(0)=0\,,\quad \text{and}\quad c_q'(0)=0\,.
\ee 
These IC describe a kink with the shape mode excited with an initial amplitude of $A_0$ and no radiation. Due to the addition of the dissipative degrees of freedom (radiation modes), the effective system (\ref{effec_eq:cq})-(\ref{effec_eq:cs}) naturally allows the decay of the shape mode. In Fig. \ref{fig:manton} we compare the decay of the shape mode obtained from field theory with the solution of the effective system (\ref{effec_eq:cq})-(\ref{effec_eq:cs}). Notice that, the bigger the number of scattering modes added to the effective model, the better the fit to field theory. We see that, for times $t_c<1/\Delta q$ and number of scattering modes $N>5$, both solutions agree with great accuracy. 

As we have learnt from the previous section, the most relevant scattering modes to describe the decay should have frequencies close to $\omega=2\omega_s$, since these modes carry the energy to infinity. In terms of the system  (\ref{effec_eq:cq})-(\ref{effec_eq:cs}) this has a simple explanation: if only modes with frequency $\omega$ far from $2\omega_s$ are allowed, the forced harmonic oscillator equation (\ref{effec_eq:cq}) never enters in the resonant region $\omega=2\omega_s$, and as a result the scattering modes excited are not able to carry enough energy from the shape mode. For small shape mode amplitudes and radiation frequencies close to $2\omega_s$ the equation (\ref{effec_eq:cs}) enters in the unstable region, and the amplitude of the shape mode grows exponentially.  Assuming a monochromatic wave of the form (\ref{lin_rad}), the equation (\ref{effec_eq:cs}) reduces to
\bea\label{mathieu}
&&\ddot{c}_s(t)+\left(\omega_s^2 + f(q_0)\sin(\omega_{q_0} t)\right) c_s(t)=0\,,
\eea 
with
\bea
&& f(q_0) = - \frac{3\pi A_{q_0}}{4} \dfrac{q_0^2 (q_0^2 - 2)}{\sinh \left( \pi  q_0 / 2 \right)}\sqrt{\dfrac{q_0^2 + 4}{q_0^2 + 1}}\,. 
\eea 
The equation (\ref{mathieu}) constitutes a Mathieu equation. By means of the general theory of Mathieu equations we can compute the bands of instability. For small $A_{q_0}$ they satisfy the relation $\omega_s/\sqrt{q_0^2+4} =k/2$, where $k \in \mathbb{Z}$. For $k=1$ the instability appears when the frequency of the radiation is twice the frequency of the shape mode. Hence, the radiation triggers resonantly the shape mode and one should expect an exponential amplification of its amplitude. Due to energy conservation, as $c_s(t)$ grows, the third term in (\ref{effec_eq:cq}) transfers energy to the radiation modes and the exponential growth stops. Notice that this resonant transfer mechanism between internal modes has been also observed in two and three-dimensional solitons \cite{q1,q2}.
\begin{figure}[H]
\begin{center}
\hspace{-0.8cm}
\includegraphics[scale=0.9]{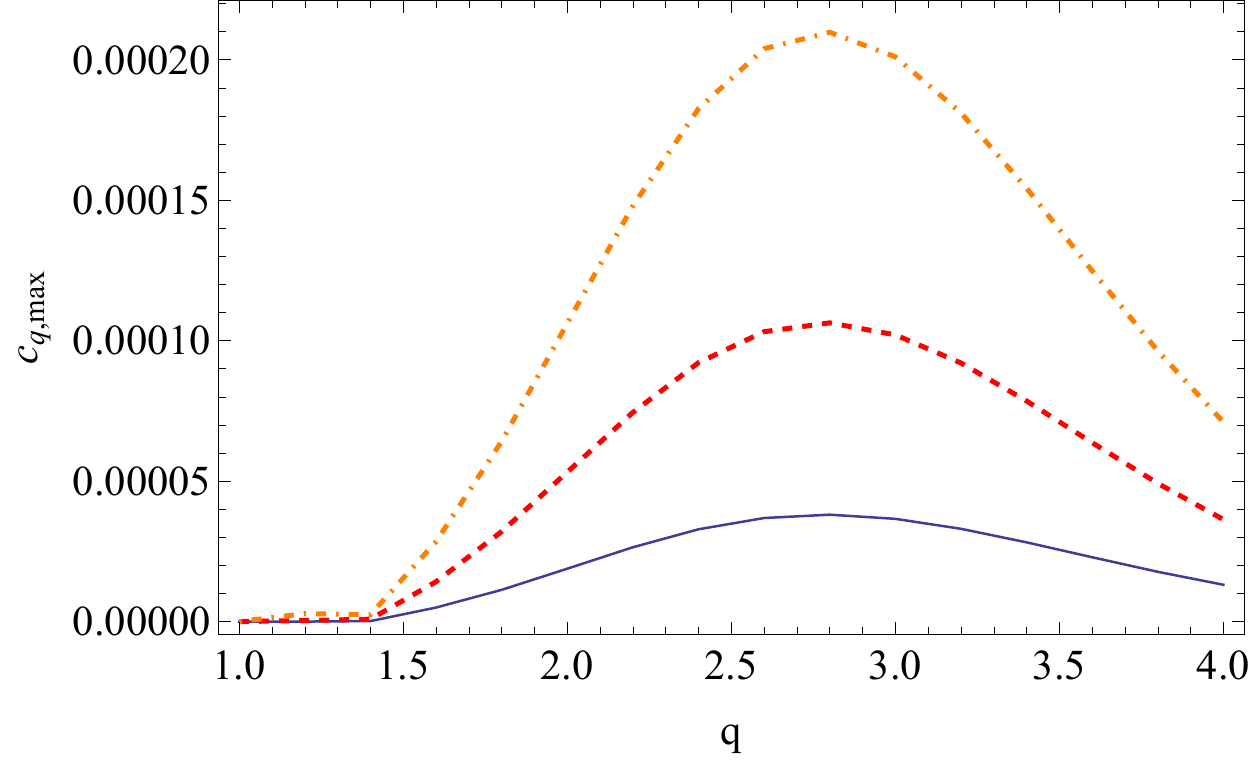}    
\caption{\small Frequency spectrum of the radiation emitted by a wobbling kink obtained from (\ref{effec_eq:cq})-(\ref{effec_eq:cs}). We have taken $n = 40$ equidistant scattering modes in the interval $q \in [-4,4]$, time of simulation $t = t_c$, and the initial values of the shape mode $c_s(0)=0.03,\, 0.05$ and $0.07$ (solid, dashed and dotted-dashed line respec.). It can be appreciated that the maxima take place approximately at $\omega_q = 2\omega_s$ for small shape mode amplitudes. }
\label{fig:excitation_rad}
\end{center}
\end{figure}
 
\begin{figure}[H]
\centering
\hspace{-0.5cm}
\subfloat[$q = 2.0$]{
\includegraphics[scale=0.635]{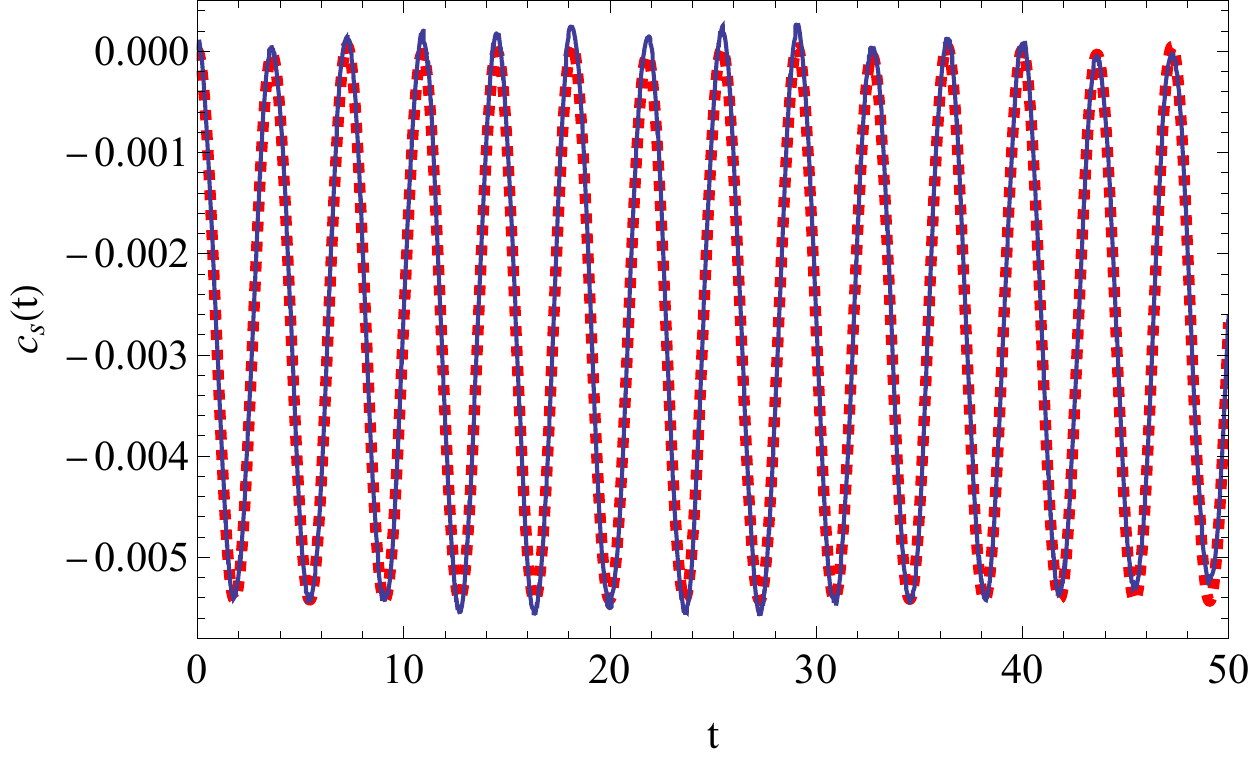} 
}
\subfloat[$q = 3.0$]{
\includegraphics[scale=0.635]{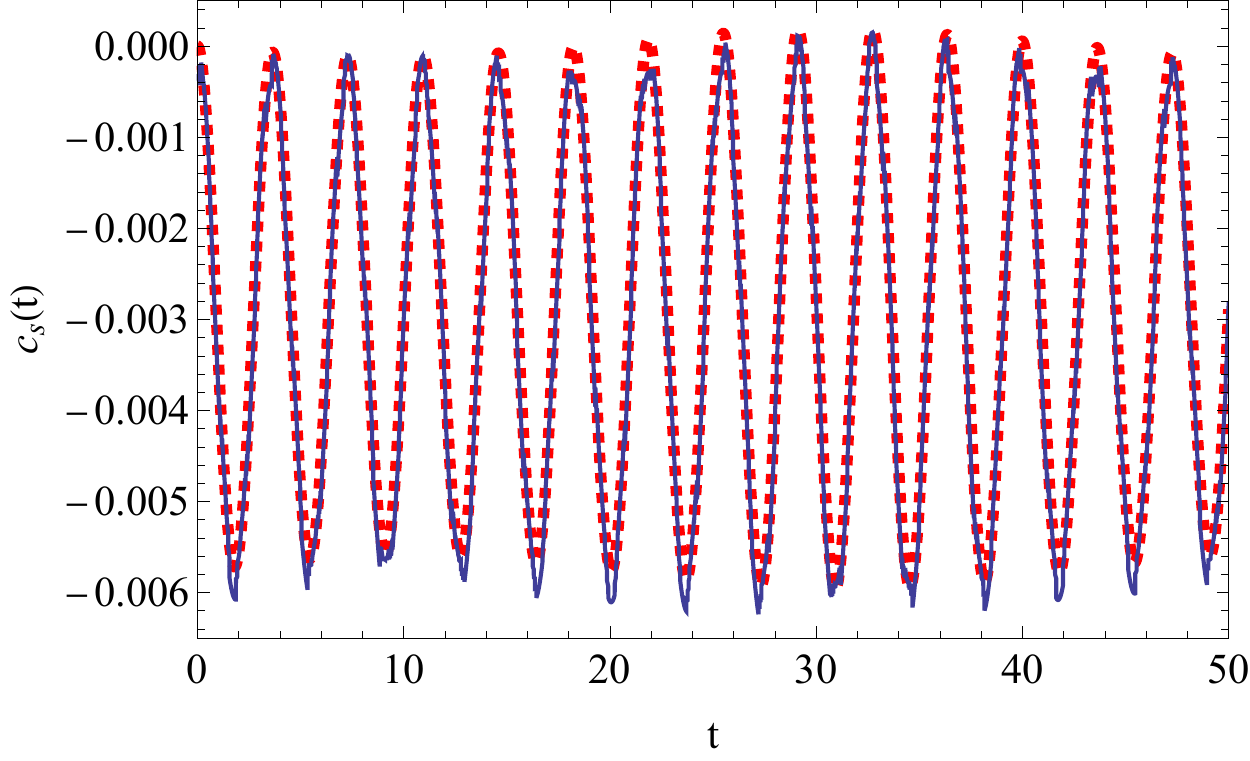}   
}
\caption{{\small Shape mode excitation by external radiation of different frequencies and amplitude $A_q = 0.02$ in field theory (solid line) and in the effective theory (dashed line) (\ref{effective_lagran:oscill}). We have taken into account $n =30$ equidistant scattering modes in the interval $q \in [-3,3]$.}}
\label{fig:excitation_shape1}
\end{figure}

In our second numerical experiment we will explore the excitation of the shape mode when the kink is illuminated with radiation of frequency $\omega_q$. We choose the following IC
\bea
\label{in_rad_1}
&&c_s(0)=0\,,\quad c_s'(0)=0\,,\\
\label{in_rad_2}
&&c_q(0)=0\,,\quad  \dot{c}_q(0)=0\,, \,\, \text{for}\,\, q\neq q_0, - q_0\,,\\\label{in_rad_3}
&&c_q(0)=A_q\,,\quad  \dot{c}_q(0)=i A_q \omega_q, \,\, \text{for}\,\, q= q_0,\\\label{in_rad_4}
&&c_q(0)=A_q\,,\quad  \dot{c}_q(0)=- i A_q \omega_q, \,\, \text{for}\,\,q= -q_0\,. 
\eea
In linear theory, the IC (\ref{in_rad_3}) and (\ref{in_rad_4}) correspond to the solution
\be
c_q(t)=A_q e^{i\omega_q t}\delta(q-q_0)+A_q e^{-i\omega_q t}\delta(q+q_0)\,.
\ee
This choice describes a superposition of a kink with a combination of scattering modes of frequency $\omega_{q_0}$. Initially, the radiation has the form
\bea\label{lin_rad}
\int_{\mathbb{R}} dq \, c_q(t) \eta_q(x) &=& \frac{A_{q_0}}{\sqrt{(q_0^2+1)(q_0^2+4)}} \left[6\tanh(x)( \cos(\omega_0 t + q_0 x)\tanh(x) + q_0 \sin (\omega_0 t + q_0 x)) \right. \nonumber\\
&&\left.-  2 (1 + q_0^2)\cos(\omega_0 t + q_0 x)\right].
\eea
Notice that, even though asymptotically (\ref{lin_rad}) is a plane wave of frequency $\omega_0$, close to the origin it gets distorted by the presence of the kink. In Fig. \ref{fig:excitation_shape1} we compare the excitation of the shape mode in field theory with the Eqs. (\ref{effec_eq:cq})-(\ref{effec_eq:cs}) with IC (\ref{in_rad_1})-(\ref{in_rad_4}). 

\begin{figure}[H]
\centering
\hspace{-0.5cm}
\subfloat[$q = 2.0$]{
\includegraphics[scale=0.635]{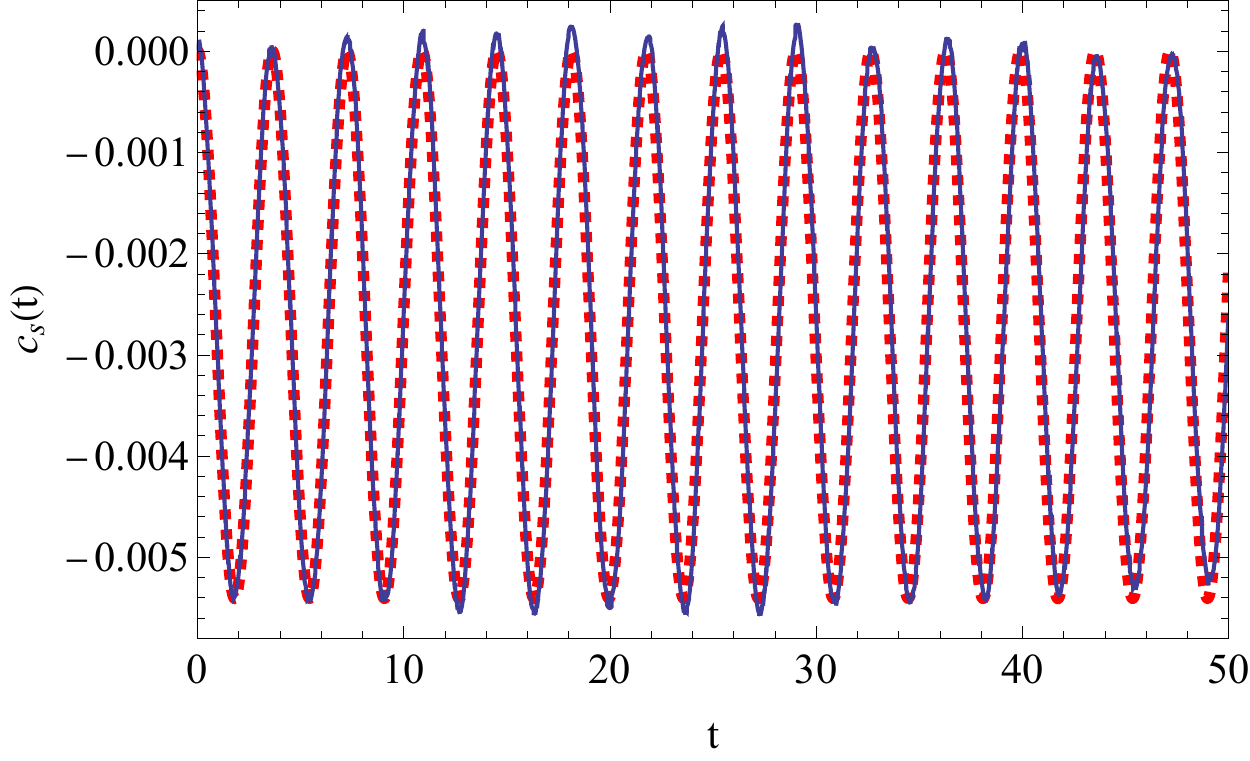}   
}
\subfloat[$q = 3.0$]{
\includegraphics[scale=0.635]{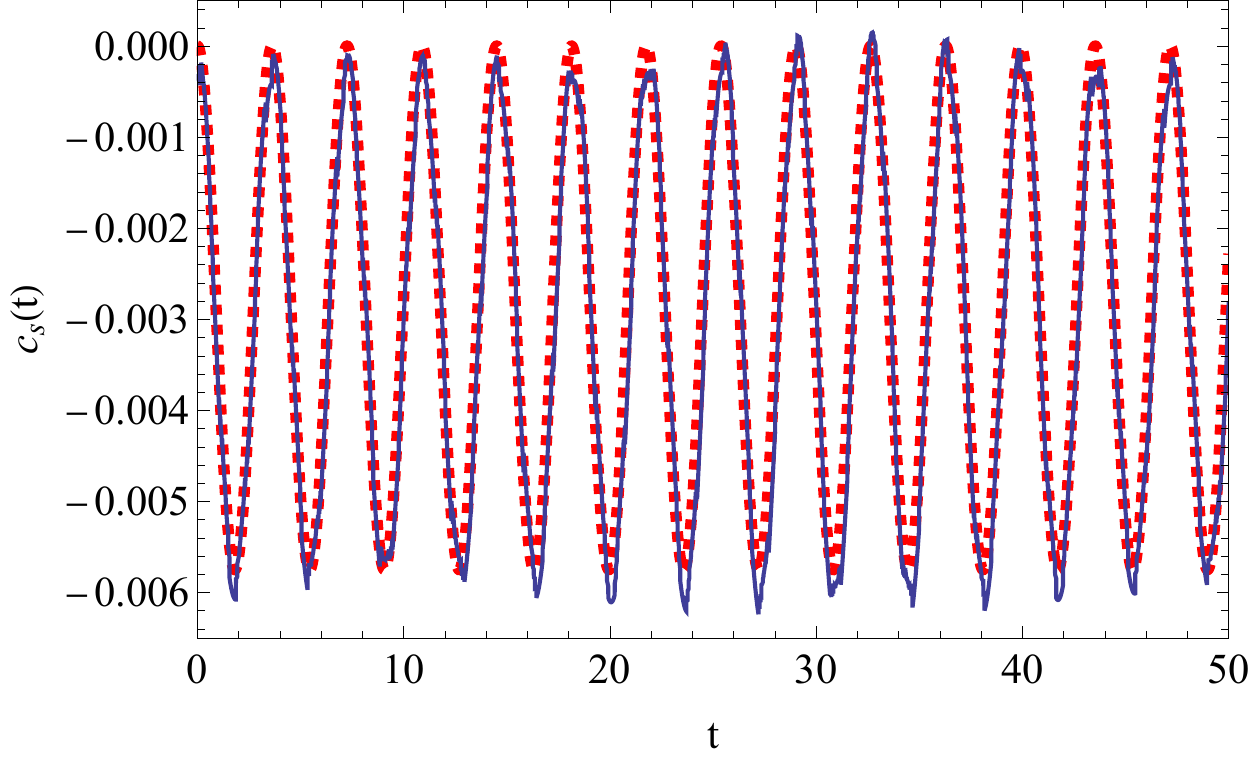} 
}
\caption{{\small Shape mode excitation by external radiation of different frequencies and amplitude $A_q = 0.02$ in field theory (solid line) and with the approximate solution (dashed line) (\ref{eq_rad_appr}). We have taken into account $n =30$ equidistant scattering modes in the interval $q \in [-3,3]$.}}
\label{fig:excitation_shape2}
\end{figure}

Quite remarkably, we can obtain an approximate analytical solution for the shape mode in the background of radiation for frequencies different from the resonance frequency. For small radiation amplitudes the relevant terms in Eq. (\ref{effec_eq:cs}) are the harmonic oscillator part and the last term (radiation source). This equation has the form of a forced harmonic oscillator. Regarding the initial conditions given by (\ref{in_rad_1})-(\ref{in_rad_4}) we get
\be\label{eq_rad_appr}
c_s(t)= A_{q_0}^2\Omega (q_0)\left(\dfrac{1}{\omega _s^2}+\dfrac{  \left(4 \omega _{q_0}^2-(\sech(\pi 
q_0)+1) \omega _s^2\right) \cos \left(t \omega_s\right)}{\omega _s^2\left(\omega_s^2-4 \omega_{q_0}^2 \right)}+\dfrac{  \sech(\pi  q_0) \cos \left(2 t
\omega _{q_0}\right)}{\omega_s^2-4 \omega_{q_0}^2}\right),
\ee
where
\be
\Omega (q_0)= - \dfrac{3 \sqrt{\dfrac{3}{2}} \pi  \left(8 q_0^4 + 34q_0^2+17\right)}{4 \left(q_0^4+5 q_0^2+4\right)}\,.
\ee
For larger radiation amplitudes this expression is not valid anymore, and new phenomena may appear (see for example \cite{Tom2} for $K\bar{K}$ creation). A comparison between the approximate analytical solution and the field theory results can be found in Fig. \ref{fig:excitation_shape2}. There we can appreciate the accuracy between both results. Indeed, this analytic expression allows us to explain the negative amplitude excitation of the shape mode.

So far we have not taken into account the translational degree of freedom in our model. However, the translational mode can interact with radiation and produce unexpected phenomena like the negative radiation pressure \cite{Roman}. We will explore some features of the interaction of the translational mode in our approach in the following section.


\section{Adding the translational mode}\label{Trans_Oscil_Radi}

In this section we aim to generalise the previous approach allowing for  translations of the kink. This new case will give rise to a richer structure and will allow for new couplings among the different collective coordinates.  Let us now consider a configuration of the form
\begin{equation}\label{ansatz:trans}
\phi (x,t) = \phi_{K}\big( x - a(t) \big) + c_{s}(t)\eta_{s}\big( x - a(t) \big) + \int_{\mathbb{R}} dq\, c_{q}(t)\eta_{q} \big( x - a(t) \big),
\end{equation}
where $a(t)$ is the collective coordinate that describes the translation of the kink (or the kink zero mode). Once more, we have to substitute the field configuration ansatz into the Lagrangian density of the full theory (\ref{action}). Notice that (\ref{ansatz:trans}) only adds new contributions to the kinetic part with respect to (\ref{ansatz:oscill}), whereas the potential terms remain the same. The additional contributions to the effective Lagrangian due to the presence of the zero mode are given by 
\begin{eqnarray}\label{expansion_action:kinetic_trans}
\mathcal{L}_{t} &=&\dfrac{1}{2}\int_{\mathbb{R}} dx\, \bigg( \dot{a}^{2}(t) \bigg( \int_{\mathbb{R}} dq\, dq'\, c_{q}(t)c_{q'}(t)\eta'_{q}(x)\eta'_{q'}(x) + \phi'\hspace{0.01cm}_{K}^{2}(x) + c_{s}^{2}(t)\eta'\hspace{0.01cm}_{s}^{2}(x) \nonumber\\
\!\!&\!\! \!\!&\!\! + 2\phi'_{K}(x)c_{s}(t)\eta'_{s}(x) + 2\phi'_{K}(x)\int_{\mathbb{R}} dq\, c_{q}(t)\eta'_{q}(x) + 2 c_{s}(t)\eta'_{s}(x)\int_{\mathbb{R}} dq\, c_{q}(t)\eta'_{q}(x)\bigg)\nonumber\\
\!\!&\!\! \!\!&\!\!  - \dot{a}(t) \bigg( 2\dot{c}_{s}(t)\eta_{s}(x)\int_{\mathbb{R}} dq\, c_{q}(t)\eta'_{q}(x) + 2c_{s}(t)\eta'_{s}(x)\int_{\mathbb{R}} dq\, \dot{c}_{q}(t)\eta_{q}(x)\nonumber\\
\!\!&\!\! \!\!&\!\! + 2\int_{\mathbb{R}} \,dq \,dq'\, c_{q'}(t)\dot{c}_{q}(t)\eta_q(x)\eta'_{q'}(x) \bigg) \bigg).
\end{eqnarray}

Integrating in the $x-$coordinate, adding $\mathcal{L}_{s,q}$, and assuming $\abs{\dot{a}(t)} \ll 1$, we finally get the following effective Lagrangian
\bea\label{Lag_sqt}
\mathcal{L}_{s,q,t}&=& \dfrac{1}{2}\left(\dot{c}^2_s(t)-\omega_s^2 c_s^2(t)\right) + \pi \int dq \left(\dot{c}_{q}(t)\dot{c}_{-q}(t) - \omega_q^2 c_{q}(t)c_{-q}(t)\right) + c_s^2(t)\int dq f_s(q)c_q(t)\nonumber\\
&+& c_s(t)\int dqdq' f_{sq}(q,q')c_q(t)c_{q'}(t) + \dfrac{2}{3}\dot{a}^2(t) + \dfrac{\pi}{4}\sqrt{\dfrac{3}{2}}\dot{a}^2(t)c_s(t)+\dot{a}^2(t)\int dq f_{aa}(q)c_q(t) \nonumber\\
&+& \dot{a}(t)\int dq f_{as}(q)\left(\dot{c}_s(t)c_q(t)-c_s(t)\dot{c}_q(t)\right) + \dot{a}(t)\int dq dq' f_a(q,q') \dot{c}_{q}c_{q'}(t).
\eea
The couplings between modes are collected in the following functions
\bea
f_s(q)&=& - \frac{3 i \pi}{16}\sqrt{\frac{q^2+4}{q^2+1}}\dfrac{q^2(q^2-2)}{\sinh\left(\pi q / 2\right)},\\
f_{as}(q)&=& - \dfrac{\pi}{4}\sqrt{\frac{3}{2}}\sqrt{\frac{q^2+1}{q^2+4}}\dfrac{q^2 + 3}{\cosh\left(\frac{\pi q}{2}\right)}\,,\\
f_{aa}(q)&=& -i \dfrac{\pi}{4}\sqrt{\dfrac{q^2+4}{q^2+1}}\dfrac{q^2}{\sinh\left(\pi q/2\right)}\,,\\
f_a(q,q')&=&\dfrac{3 i \pi}{4}\frac{\left(4+q^2+q'^2\right)}{\sqrt{(q^2+1)(q^2+4)}\sqrt{(q'^2+1)(q'^2+4)}}\dfrac{q^2 - q'^2}{\sinh(\dfrac{\pi}{2}(q + q'))}\nonumber\\
& & - \dfrac{2 i\pi q' \left(4-9 q q' - 2 q^2 - 2 q'^2 + q^2 q'^2\right)}{\sqrt{(q^2+1)(q^2+4)}\sqrt{(q'^2+1)(q'^2+4)}}\delta(q+q')\,,\\
f_{sq}(q,q')&=&-\frac{3\pi}{8}\sqrt{\frac{3}{2}}\frac{17+17q^2+17 q'^2+10 q^2q'^2-q^4-q'^4+q^2q'^4+q^4q'^2-q^6-q'^6}{\sqrt{(q^2+1)(q^2+4)}\sqrt{(q'^2+1)(q'^2+4)}\cosh(\frac{\pi}{2}(q+q'))}.
\eea

The quadratic terms in (\ref{Lag_sqt}) are just a collection of harmonic oscillators plus the kinetic term of the kink. On the other hand, the cubic terms describe the interaction between the modes. The field equations associated to (\ref{Lag_sqt}) for the radiation modes are yielded by
\bea\label{eq_rad_kink}
 \ddot{c}_{-q}(t) &+& \omega_{q}^2 c_{-q}(t) - \dfrac{1}{2\pi}c^2_s(t) f_s(q) - \dfrac{1}{\pi} c_s(t)\int dq' f_{sq}(q,q')c_{q'}(t) - \dfrac{1}{2\pi}\dot{a}^2(t)f_{aa}(q)\nonumber\\
&-& \dfrac{1}{2\pi}\ddot{a}(t)f_{as}(q)c_s (t) - \dfrac{1}{\pi}\dot{a}(t)f_{as}(q)\dot{c}_s(t) +\dfrac{1}{2\pi}\dot{a}(t)\int dq'  \dot{c}_{q'}(t)\left(f_a(q,q')-f_a(q',q)\right)\nonumber\\
&+& \dfrac{1}{2\pi}\ddot{a}(t)\int dq' f_a(q,q')c_{q'}(t) = 0,
\eea
whilst the equations of motion determining the evolution of $c_s(t)$ and $a(t)$ take the form
\bea\label{eq_shape_kink}
\ddot{c}_s(t) &+& \omega_s^2 c_s(t) - 2 c_s(t)\int dq f_s(q)c_q(t) - \int dqdq' f_{sq}(q,q')c_q(t)c_{q'}(t) - \frac{\pi}{4}\sqrt{\frac{3}{2}}\dot{a}^2(t) \nonumber \\ 
& +& 2\dot{a}(t)\int dq  f_{as}(q)\dot{c}_{q}(t) + \ddot{a}(t)\int dq f_{as}(q)c_q(t) = 0\,,
\eea
and
\bea\label{eq_trans_kink}
\frac{4}{3}\ddot{a}(t)&+&\dfrac{\pi}{2}\sqrt{\dfrac{3}{2}}\left(\ddot{a}(t)c_s(t) + \dot{a}(t)\dot{c}_s(t)\right) + 2\int dq f_{aa}(q)\left(\ddot{a}(t)c_q(t) + \dot{a}(t)\dot{c}_q(t)\right)  \nonumber\\
 &+&\int dq f_{as}(q)\left(\ddot{c}_s(t)c_q(t) - c_s(t)\ddot{c}_q(t)\right) + \frac{d}{dt}\int dq dq' f_a(q,q') \dot{c}_{q}c_{q'}(t) = 0\,.
\eea

The first configuration we will discuss is the translating kink without any excitation. In the non-relativistic approach this corresponds to $a(t)=x_0 + v t$, while the rest of the modes vanish. However, a simple inspection of the system (\ref{eq_rad_kink})-(\ref{eq_trans_kink}) reveals something surprising at first glance: if $\dot{a}(t)\neq0$, the terms proportional to $\dot{a}(t)^2$ in (\ref{eq_rad_kink}) and (\ref{eq_shape_kink}) act as sources for the shape mode and radiation. This seems contradictory since, due to the Lorentz invariance of the model, the boosted kink is an exact solution. The Lorentz boosted version of the kink has the following form
\be\label{boosted}
\phi(x,t)=\tanh\left(\frac{x-v t}{\sqrt{1-v^2}}\right).
\ee
Of course, (\ref{boosted}) does not contain any radiation. This apparent pathology of the effective model can be easily resolved. The standard CCM approach is not Lorentz invariant at first order. This is obvious from the asymmetry  between time and  spatial coordinates in the ansatz (\ref{ansatz:trans}). However, once we consider higher order terms, approximate Lorentz invariant solutions exist. Actually, the mentioned source terms for $c_s(t)$ and $c_q(t)$ are responsible for the Lorentz contraction of a moving kink. Let us consider a moving kink with velocity $v<\!\!<1$ located at the origin, then initially $\dot{a}(t)=v$. Assuming that $c_s(t)$ and $c_q(t)$ do not depend on time, an approximate solution of (\ref{eq_shape_kink}) and (\ref{eq_rad_kink}) is given by
\be\label{sol_boost}
c_s(t)=\frac{\pi }{4\sqrt{6}}v^2,\quad c_q(t)=-\frac{i q^2\csch(\pi q/2)}{8\sqrt{(q^2+1)(q^2+4)}}v^2,
\ee
where we disregard corrections of order $\mathcal{O}(v^4)$. On the other hand, if we expand (\ref{boosted}) with respect to $v$, at $t=0$ we get
\be
\phi(0,x)=\tanh(x)+\frac{1}{2}\left(x-x\tanh^2(x)\right)v^2+\mathcal{O}(v^4).
\ee
We denote the first correction to the Lorentz contraction by $\phi^{(1)}(x)=\frac{1}{2}\left(x-x\tanh^2(x)\right)v^2$. The projection of $\phi^{(1)}(x)$ onto the spectral modes gives
\bea
\langle \phi^{(1)}(x),\eta_s(x) \rangle&=&\frac{\pi }{4\sqrt{6}}v^2\,, \\
\langle \phi^{(1)}(x),\eta_q(x) \rangle&=&-\frac{i \pi q^2\csch(\pi q/2)}{4\sqrt{(q^2+1)(q^2+4)}}v^2\,.
\eea

Note that there is an extra  $2\pi$ factor due to the normalization of the scattering modes. This is easily interpreted: this effective model already describes relativistic effects at this order. The solution (\ref{sol_boost}) can be interpreted as a first order Lorentz boost, i.e., in terms of the coordinates of our model the Lorentz boosted kink takes the form
\bea
\phi_B(x,t)&=&\tanh(x- v t)+\phi^{(1)}(x-v t)+\mathcal{O}(v^4)=\nonumber \\
&=&\tanh(x-v t)+v^2\frac{\pi }{4\sqrt{6}} \eta_s(x-v t)-v^2\int dq \frac{i q^2\csch(\pi q/2)}{8\sqrt{(q^2+1)(q^2+4)}}\eta_q(x-v t),
\eea
for $x_0 = 0$. If one is interested in the description of relativistic processes, this approach does not seem appropriate since, even to describe a simple boosted solution, one needs to excite a large number of modes. Still, we would like to emphasise that, already at quadratic order, the ``non-relativistic" CCM approach describes relativistic effects, although these effects require in general, the excitation of scattering modes. There is a simple ansatz that describes the exact Lorentz boost with only two degrees of freedom given by the following expression
\be\label{rel_ansatz}
\phi(x,t)=\phi_K\left(\frac{x-a(t)}{\delta(t)}\right)\,.
\ee
This type of ansatz was first considered in \cite{Rice} (for recent discussions see \cite{Relativistic}). However, after the introduction of the scattering modes, it seems not possible to obtain analytical results, thus we will not pursue this line in this section. However, as we will see in Sec. \ref{oscillon}, some small generalizations of (\ref{rel_ansatz}) may be used to describe effectively dissipative degrees of freedom. 

\begin{figure}[H]
\begin{center}
\hspace{-0.8cm}
\includegraphics[scale=0.9]{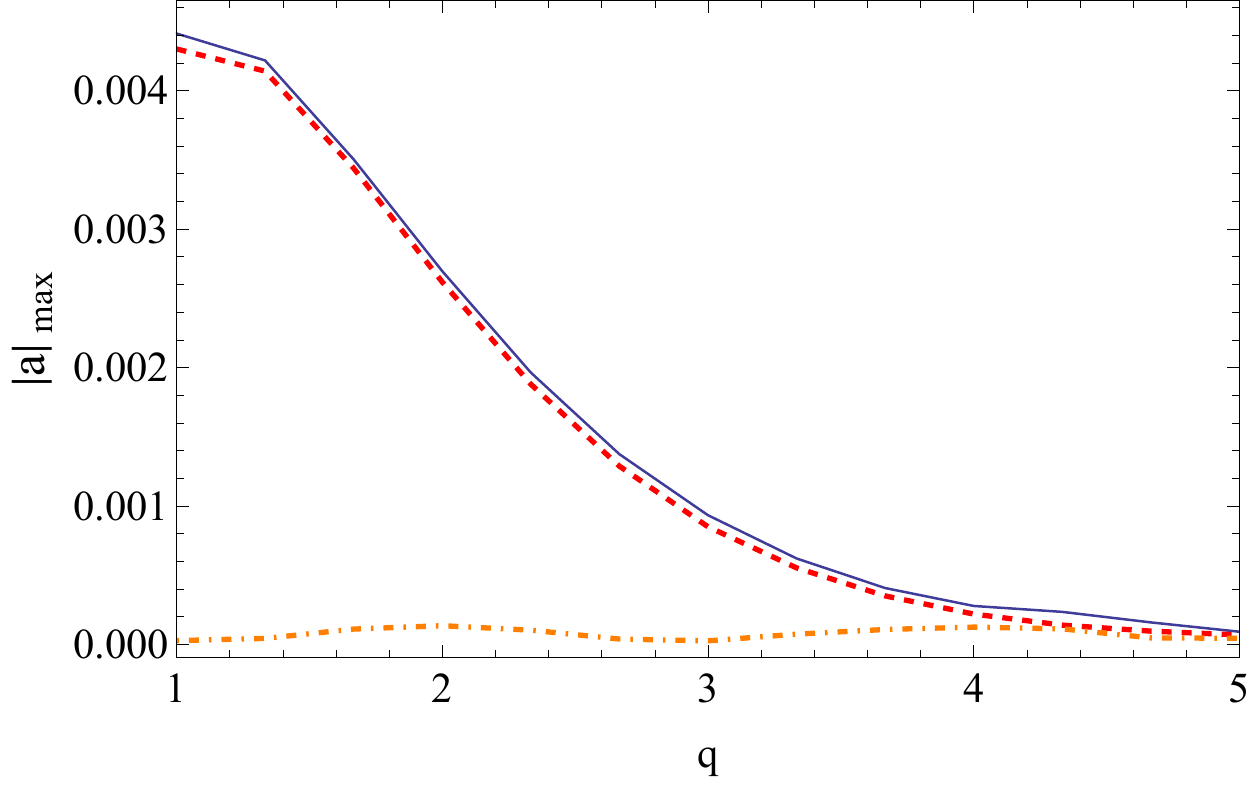}    
\caption{\small Maximum displacement of the kink due to radiation of different frequencies and amplitude $A_q = 0.05$. The solid line represents the complete effective model, the dashed line represents the simulation when only the $f_{as}$ source is considered, and the dotted dashed represents the simulation when only the source $f_{a}$ is taken into account. We have taken into account $n = 30$ equidistant scattering modes in the interval $q \in [-5,5]$. The time of simulation is $t = t_c$.}
\label{fig:displacement}
\end{center}
\end{figure}

In our second experiment we will study the excitation of the translational mode by radiation. Notice that the first and last term of (\ref{eq_trans_kink}) resemble the Newton equation, where we can identify $m_K = 4/3$. However, for the IC representing a kink at rest illuminated by linear radiation (\ref{lin_rad}), the rate of change of the total momentum of radiation given by the last term in (\ref{eq_trans_kink}) vanishes, so the kink remains at rest. As a matter of fact, it is well-known that the $\phi^4$ kink is transparent to linear radiation as it was emphasised in \cite{Roman}, and that the negative radiation pressure appears at fourth order in perturbation theory. Nevertheless, the shape mode may act as an intermediary between the zero mode and radiation. Namely, a non-trivial interplay between the modes could be as follows: the radiation triggers the shape mode due to the term proportional $f_{sq}(q,q')$ in (\ref{eq_shape_kink}). Once $c_s(t)$ is excited,  there is a source for $a(t)$ in (\ref{eq_trans_kink}) given by the term proportional to $f_{as}(q)$. The excitation of $a(t)$ is roughly of order $A_0A_q$, where $A_0$ is the amplitude of the shape mode and $A_q$ is the amplitude of the $q-$scattering mode. Since from (\ref{eq_rad_appr}) $A_0$ is of order $A_q^2$, the excitation of the zero mode due to this mechanism appears at order $\mathcal{O}(A_q^3)$. However, contrary to the negative radiation pressure effect, there is no net momentum transfer at this order and the kink simply oscillates about its rest position. This can be seen easily from (\ref{eq_trans_kink}) disregarding the last term and keeping terms of $\mathcal{O}(A_q^3)$ order. In Fig. \ref{fig:displacement} we show the maximum amplitude of the zero mode as a function of the incident radiation frequency. This shows that for small $q$ the term proportional to $f_a$ is indeed subleading. 
\vspace{0.3cm}

For larger times, higher order corrections to radiation start to play a role and the last term in (\ref{eq_trans_kink}) cannot be neglected anymore. This term transfers a net momentum to the kink \cite{Roman} and it is pulled back in the opposite direction of the incoming radiation. 

So far, we have discussed only the single-kink sector. Of course, one expects that the combination of translational and scattering modes should play an important role in the $K\bar{K}$ sector, allowing for energy dissipation in scattering processes or even describing $K\bar{K}$ annihilation. This is a more ambitious goal that we leave for future research. However, during a kink-antikink  scattering process, when the solitons completely overlap, an intermediate state that resembles the profile of an oscillon is formed. This suggest that the study of oscillons, with appropriate initial conditions, may provide useful information about these violent processes. This is the point of view we adopt in the following section.

 
\section{Effective model for the radiating oscillon}
\label{oscillon}

\begin{figure}[H]
\centering
\subfloat[$n = 20$, $a_0 = 0.3$, $R = 1.5.$]{
\includegraphics[scale=0.6]{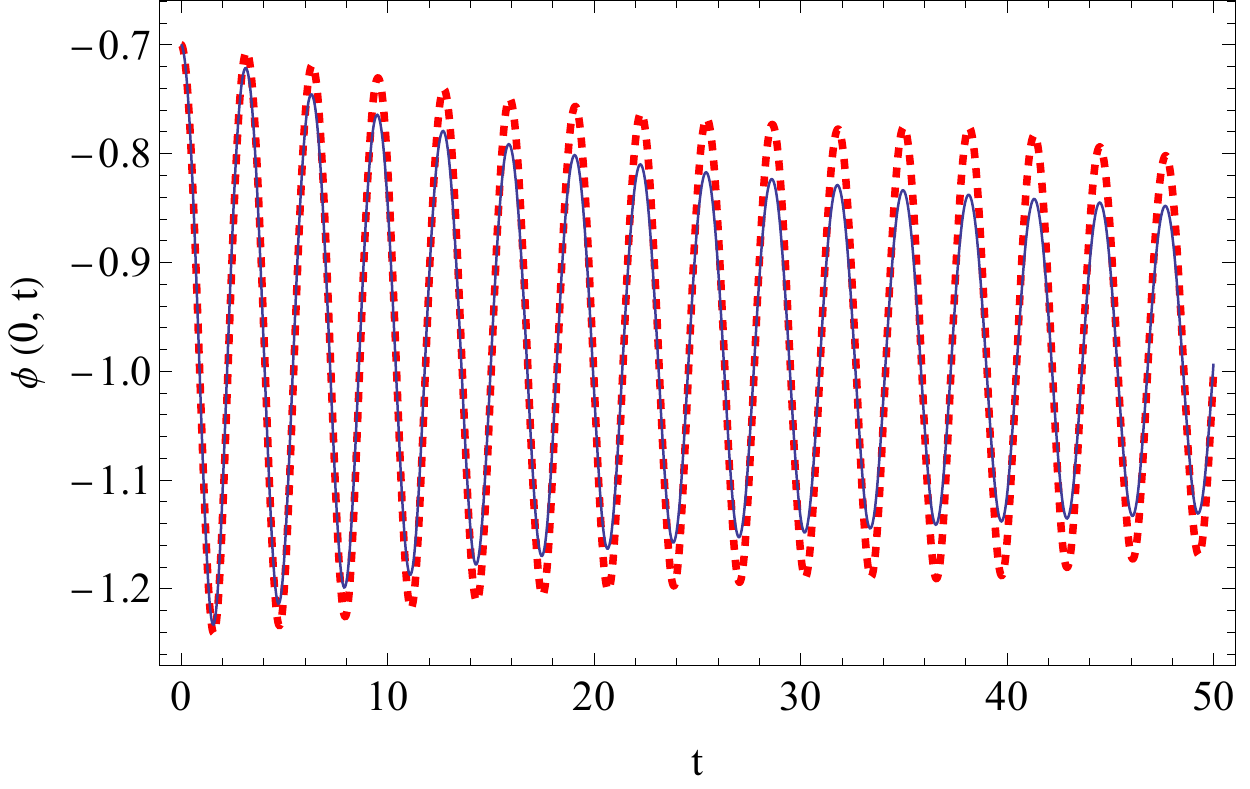}   
}
\subfloat[$n = 30$, $a_0 = 0.2$, $R = 1.5.$]{
\includegraphics[scale=0.6]{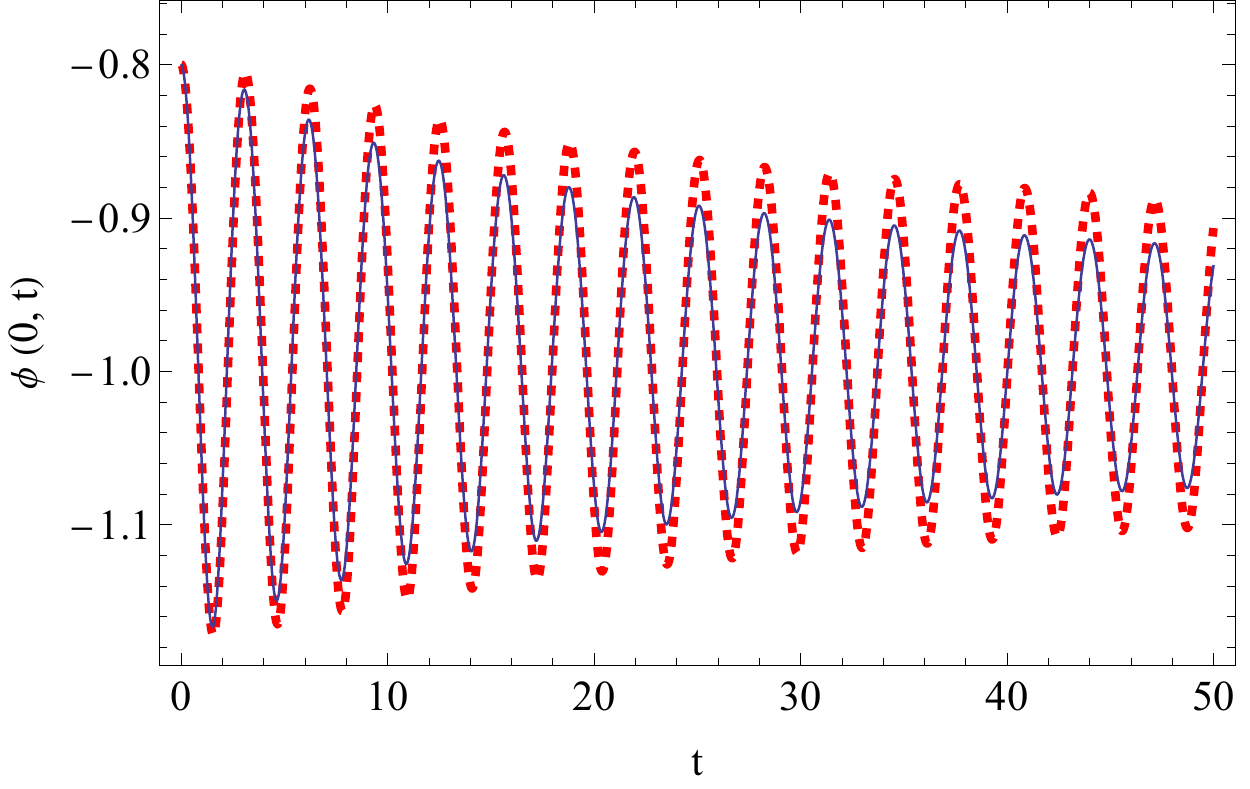}   
}\\
\subfloat[$n = 10$, $a_0 = 0.4$, $R = 2.$]{
\includegraphics[scale=0.6]{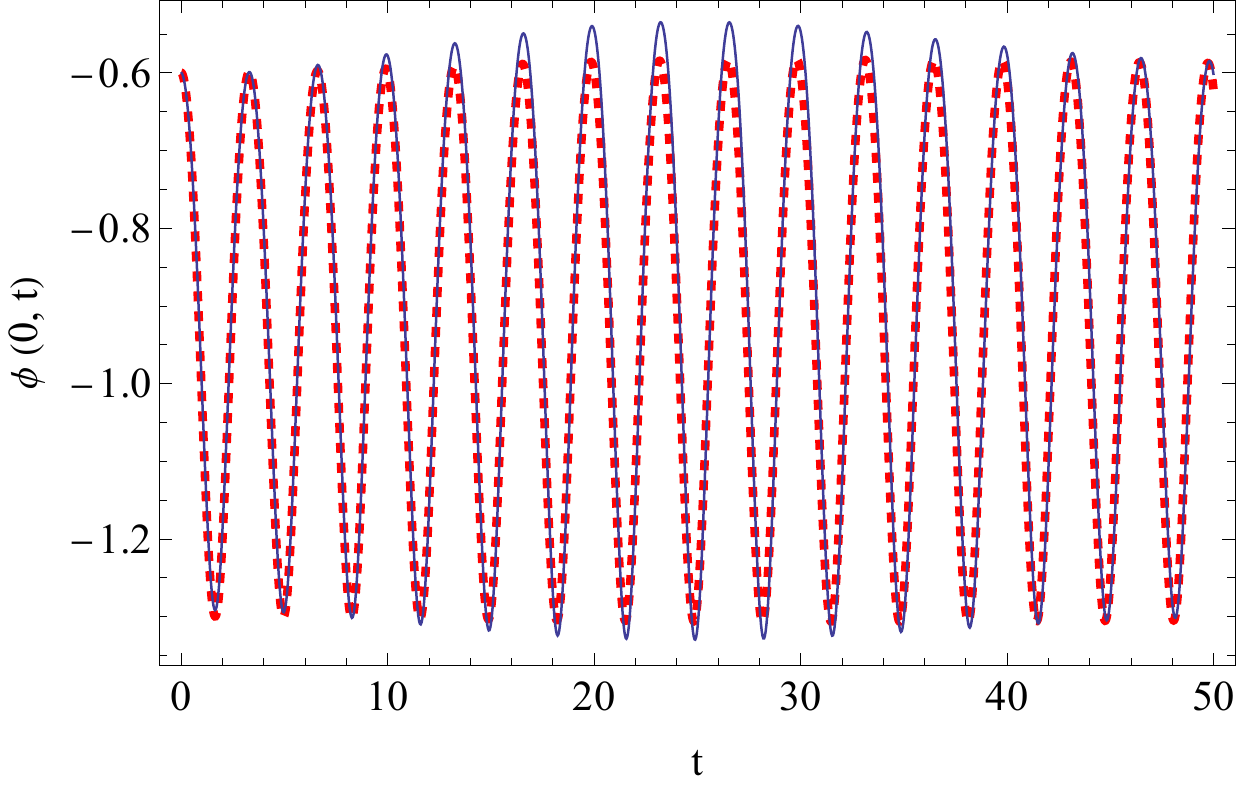}   
}
\subfloat[$n = 10$, $a_0 = 0.5$, $R = 2.$]{
\includegraphics[scale=0.6]{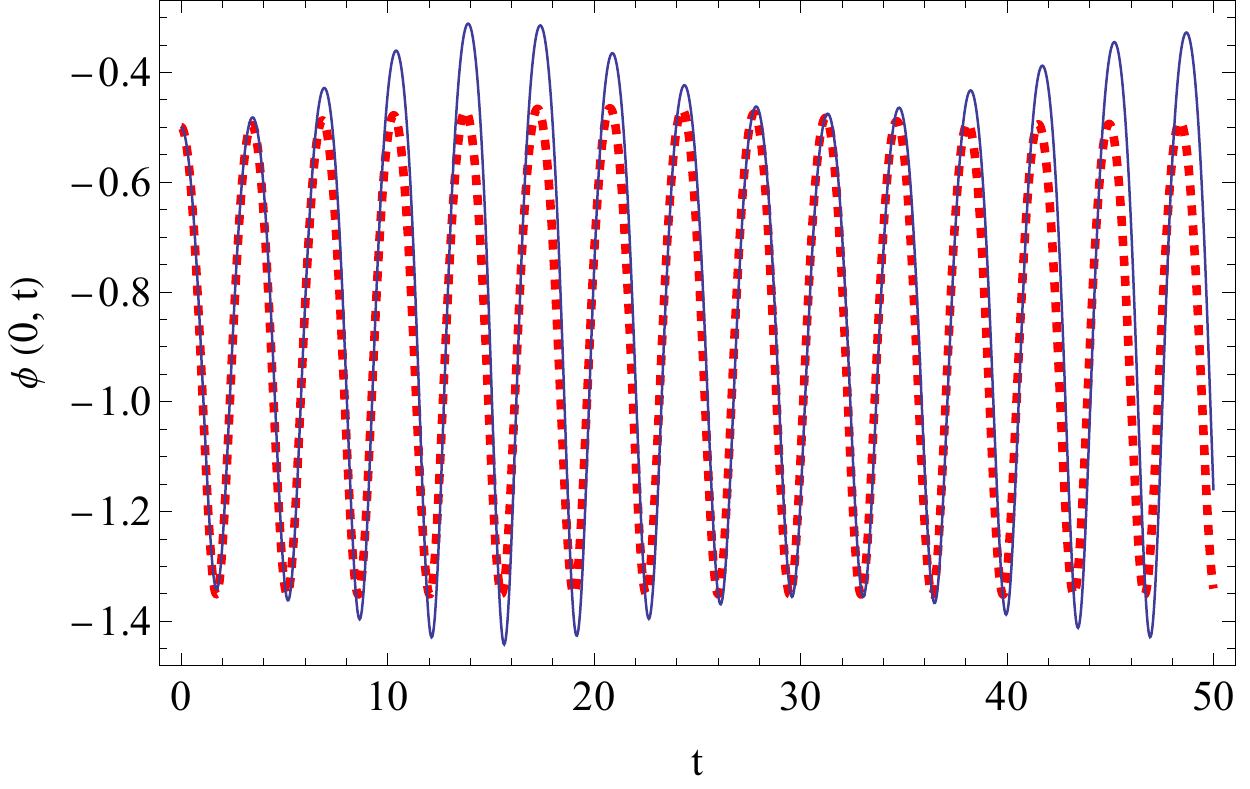}    
}
\caption{\small Comparison between the effective model (dashed line) (\ref{lag_oscill_rad}) and field theory (solid line). The scattering modes have been taken in the interval $q \in [-5,5]$.}
\label{oscillon_radiation}
\end{figure}

As we have shown, the scattering modes play an important role in the dynamics of the single-kink sector. In this section we will illustrate the importance of such modes for a non-topological soliton, the oscillon of the 1-dimensional $\phi^4$ model. In particular, we will describe the decay of oscillons below the critical amplitude, the possibility of internal modes and the $K\bar{K}$ formation. Since oscillons have been observed starting from rather generic initial data, we decide to follow first \cite{Fodor1, Fodor2}, and take the following ansatz
\be\label{KAK_ansazt_1}
\Phi_\text{o}(x;a)= -1 + a\sech(x/R), 
\ee
where $a$ is identified with the amplitude of the oscillon and $R$ accounts for its size. Later, in this section, we will change to a Gaussian profile used in \cite{Hindmarsh}. The effective Lagrangian associated to (\ref{KAK_ansazt_1}) assuming that $a$ depends on time takes the following form
\be\label{Lag_eff_osci}
\mathcal{L}^{o}= R\left(\dot{a}^2(t)-\dfrac{1}{3}(12+\dfrac{1}{R^2})a^2(t)+\pi a^3(t)-\dfrac{2}{3}a^4(t)\right),
\ee
This is the Lagrangian of an anharmonic oscillator of frequency $\omega_o = \sqrt{\frac{1}{3}(12-\frac{1}{R^2})}$. For small amplitudes, the frequency of $a(t)$ is above the threshold frequency $\omega_t = 2$. As a result, $a(t)$ couples directly to the continuum and collapses into radiation. As a consequence, for $a$ small enough, the initial data (\ref{KAK_ansazt_1}) does not evolve into an oscillon.  However, for $a(t)$ large enough, the non-linearities in (\ref{Lag_eff_osci}) decrease the oscillator frequency avoiding the direct coupling to radiation. Therefore, the oscillon weakly couples to radiation and its amplitude decreases very slowly. The details of this coupling are, of course, not described in (\ref{Lag_eff_osci}), and one needs to add dissipative degrees of freedom. This can be done as in the previous sections considering an ansatz of the form
\be\label{KAK_ansazt_rad}
\Phi_\text{o, rad}(x;a,c_q)= - 1 + a(t) \sech(x/R)+\int_\mathbb{R} dq c_q(t)\eta_q(x/R).
\ee
It is enough to consider the truncated Lagrangian at second order in $c_q(t)$ since we are interested in the description of the oscillon coupled to the slow-amplitude emitted radiation. However, as we have mentioned, it is fundamental to consider the action at all orders in $a(t)$, since the existence of the oscillon is linked to the non-linear structure of the Lagrangian. Proceeding as in the previous sections we arrive at
\begin{eqnarray}\label{lag_oscill_rad}
\mathcal{L}^{o}_{r} &=& \pi \left( \int_{\mathbb{R}} dq\, \dot{c}_q(t)\dot{c}_{-q}(t) - \int_{\mathbb{R}} dq\, w_{q,R}^2\, c_q(t)c_{-q}(t) \right) + \dot{a}^{2}(t) - \omega_o^2 a^{2}(t) + \pi a^{3}(t) - \dfrac{2}{3}a^{4}(t)\nonumber\\
& &  + \int_{\mathbb{R}^2} dq dq'\, f_1(q,q',R)\, c_{q}(t)c_{q'}(t) + \dot{a}(t)\int_{\mathbb{R}} dq\, f_2(q)\dot{c}_q(t) + a(t)\int_{\mathbb{R}} dq\, f_3(q)c_q(t)\nonumber\\
& &  + a(t)\int_{\mathbb{R}^2} dqdq'\, f_4(q,q')c_q(t)c_{q'}(t) + a^2(t)\int_{\mathbb{R}} dq dq'\, f_{5}(q,q')c_q(t)c_{q'}(t) \nonumber\\
& & + a^{3}(t)\int_{\mathbb{R}} dq\, f_6(q)c_q(t),
\end{eqnarray}
with $w_{q, R} = \sqrt{ \left(q/R\right)^2 + 4}$ and where
\begin{eqnarray}
f_1(q, q', R) &=& - \dfrac{3 \pi}{5 R^2} \dfrac{ 4 q + 4 q' + 5 q^3  + 5 q'^3  + q^5 + q'^5 }{\sqrt{\left(q^2+1\right) \left(q^2+4\right)}\sqrt{\left(q'^2+1\right) \left(q'^2+4\right)}} \csch\left(\dfrac{\pi}{2}(q + q')\right),\\
f_2(q) &=& \dfrac{\pi}{2}\dfrac{\sqrt{q^2 + 1}}{\sqrt{q^2 + 4}} \sech\left(\dfrac{\pi 
q}{2}\right) ,\\
f_3(q, R) &=& \pi\dfrac{\sqrt{q^2 + 1}}{\sqrt{q^2 + 4}}\left( \dfrac{q^2 + 3}{4 R^2} - 2\right)\sech\left(\dfrac{\pi q}{2}\right),\\
f_4(q, q') &=& \dfrac{3\pi}{4}\dfrac{11 + 14q^2 + 14q'^2 + 2q^2q'^2 + 3q^4 + 3q'^4}{\sqrt{\left(q^2+1\right) \left(q^2+4\right)}\sqrt{\left(q'^2+1\right) \left(q'^2+4\right)}}\sech\left(\dfrac{\pi}{2} (q + q')\right),\\
f_5(q, q') &=& - \dfrac{3 \pi}{5}  \dfrac{ 4 q + 4 q' + 5 q^3 + 5 q'^3 + q^5 + q'^5 }{\sqrt{\left(q^2+1\right) \left(q^2+4\right)}\sqrt{\left(q'^2+1\right) \left(q'^2+4\right)}} \csch\left(\dfrac{\pi}{2}(q + q')\right) ,\\
f_6(q) &=& \dfrac{\pi}{4}\dfrac{(q^2 + 1)^{3/2}}{\sqrt{q^2 + 4}} \sech\left(\dfrac{\pi 
q}{2}\right).
\end{eqnarray}
Notice that we have omitted a global factor $R$ in (\ref{lag_oscill_rad}) since it does not have any effect on the equations of motion. In Fig. \ref{oscillon_radiation} we illustrate the evolution of the oscillon profile at $x = 0$ for different initial amplitudes and sizes.
In the upper panel we show two configurations with small initial amplitudes. This results into a fast decay of the initial configuration into radiation. It is worth mentioning even with a large number of radiation modes ($n>20$) the amplitude in the effective model decays slower than in the field theoretical simulation. This suggests that the dissipative mechanism provided by the set of scattering modes is actually not very efficient. In the lower panel we show a genuine oscillon. The numerical simulation reveals that the oscillon hosts an internal non-dissipative mode,  due to the amplitude modulation during the time evolution. This mode can be related to the variation of the oscillon size. Hence, let us consider a small variation of the oscillon width as follows
\be\label{KAK_ansazt_size}
\Phi_\text{o}(x;a,\delta)= -1+ a\sech\left(\frac{x}{R+\delta}\right).
\ee
If $\delta \ll 1$ we may expand about $\delta=0$ up to first order 
\be\label{KAK_ansazt_size_1}
\Phi_\text{o}(x;a,\delta)=-1+a \sech(x/R)+\frac{a \delta}{R^2} \,x\, \sech(x/R)\tanh(x/R)\,.
\ee
The additional term to the unperturbed oscillon corresponds to the so-called Derrick mode. This correction should codify the possible changes of the oscillon size, and luckily represent the behaviour expected from the full numerical result. This simple choice has a problem. At $a = 0$, the ansatz (\ref{KAK_ansazt_size}) does not depend on $\delta$. This implies that the moduli metric associated to $(a, \delta)$ is not well-defined at this point (see for example \cite{Manton1} for a discussion about the null vector problem). In order to cure this issue we may perform a simple change of coordinates $\delta\rightarrow \delta/a $. Finally, we get
\be\label{KAK_ansazt_size_2}
\Phi_\text{o, s}(x;a,\delta)=-1+ a\sech(x/R)+\frac{\delta}{R^2}\, x\, \sech(x/R)\tanh(x/R)\,.
\ee
Treating $a$ and $\delta$ as collective coordinates (recall that $R$ remains fixed) we get the below effective Lagrangian
\bea\label{lag_osc_internal}
\mathcal{L}^{o}_{s} &=& R \dot{a}(t)^2 - \omega_o^2 a(t)^2 + \dfrac{1}{R}\left(\dfrac{\pi ^2}{36} + \dfrac{1}{3} \right) \dot{\delta}(t)^2 - \dfrac{1}{R}\left( \pi ^2\left(\dfrac{1}{9} + \dfrac{7}{180 R^2} \right) + \dfrac{4}{3}  \right) \delta (t)^2 + \pi  R a(t)^3\nonumber\\
& & - \dfrac{2}{3} R a(t)^4 + \dfrac{\pi}{R^2}\left( \dfrac{11 \pi ^2}{80} - 1 \right) \delta (t)^3 - \dfrac{1}{R^3}\left(\dfrac{\pi ^4}{600} + \dfrac{\pi ^2}{90} - \dfrac{2}{15}\right) \delta (t)^4 + \dot{a}(t) \dot{\delta }(t)\nonumber\\
&& + \left(\dfrac{1}{3 R^2}-4\right) a(t) \delta (t) + \dfrac{\pi}{R}\left(\dfrac{3 \pi ^2}{16} - 1 \right) a(t) \delta (t)^2 - \dfrac{1}{R^2}\left( \dfrac{7 \pi ^2}{90} - \dfrac{1}{3}\right) a(t) \delta (t)^3 \nonumber\\
& & + \pi  a(t)^2 \delta(t) - \dfrac{2}{3} a(t)^3 \delta (t) - \dfrac{\pi ^2}{15 R} a(t)^2 \delta (t)^2. 
\eea
The associated equations of motion are collected below
\bea
\ddot{ a}(t) &+& \dfrac{1}{3}\left(\dfrac{1}{R^2} + 12\right) a(t) - \dfrac{3}{2} \pi  a(t)^2 + \dfrac{4}{3}a(t)^3 + \dfrac{1}{2 R}\ddot{ \delta }(t) + \dfrac{\left(12 R^2 - 1\right)}{6 R^3}\delta (t) \nonumber\\
& -& \dfrac{\pi  \left(3 \pi ^2-16\right)}{32 R^2}\delta (t)^2 + \dfrac{\left(7 \pi ^2 - 30\right)}{180 R^3} \delta (t)^3 - \dfrac{\pi}{R}a(t) \delta (t) + \dfrac{\pi ^2}{15 R^2}a(t) \delta (t)^2\nonumber\\
&+& \dfrac{1}{R}a(t)^2 \delta (t) = 0\,,
\eea
and
\bea
\ddot{ \delta }(t) &+& \dfrac{\left(\pi ^2 \left(20 R^2 + 7\right) + 240 R^2 \right)}{5 \left(\pi ^2 + 12\right) R^2} \delta (t) -\dfrac{27 \pi  \left(11 \pi ^2 - 80\right)}{40 \left(\pi ^2 + 12\right) R}\delta (t)^2 +\dfrac{\left(3 \pi ^4 + 20 \pi ^2 - 240\right)}{25 \left( \pi ^2 + 12 \right) R^2}\delta (t)^3\nonumber\\
&+& \dfrac{18 R}{\pi ^2 + 12}\ddot{ a}(t) + \dfrac{\left(72 R^2-6\right)}{\left(\pi ^2 + 12\right) R}a(t) -\dfrac{18 \pi  R}{\pi ^2 + 12}a(t)^2 +\dfrac{12 R}{\pi ^2 + 12}a(t)^3 - \dfrac{9 \pi  \left(3 \pi ^2-16\right)}{4 \left(\pi ^2 + 12\right)}a(t) \delta (t)\nonumber\\
&+& \dfrac{3 \left(7 \pi ^2-30\right)}{5 \left(\pi ^2 + 12\right) R}a(t) \delta (t)^2 +\dfrac{12 \pi ^2}{5 \pi ^2 + 60}a(t)^2 \delta (t) = 0\,.
\eea

\begin{figure}[H]
\centering
\includegraphics[scale=0.9]{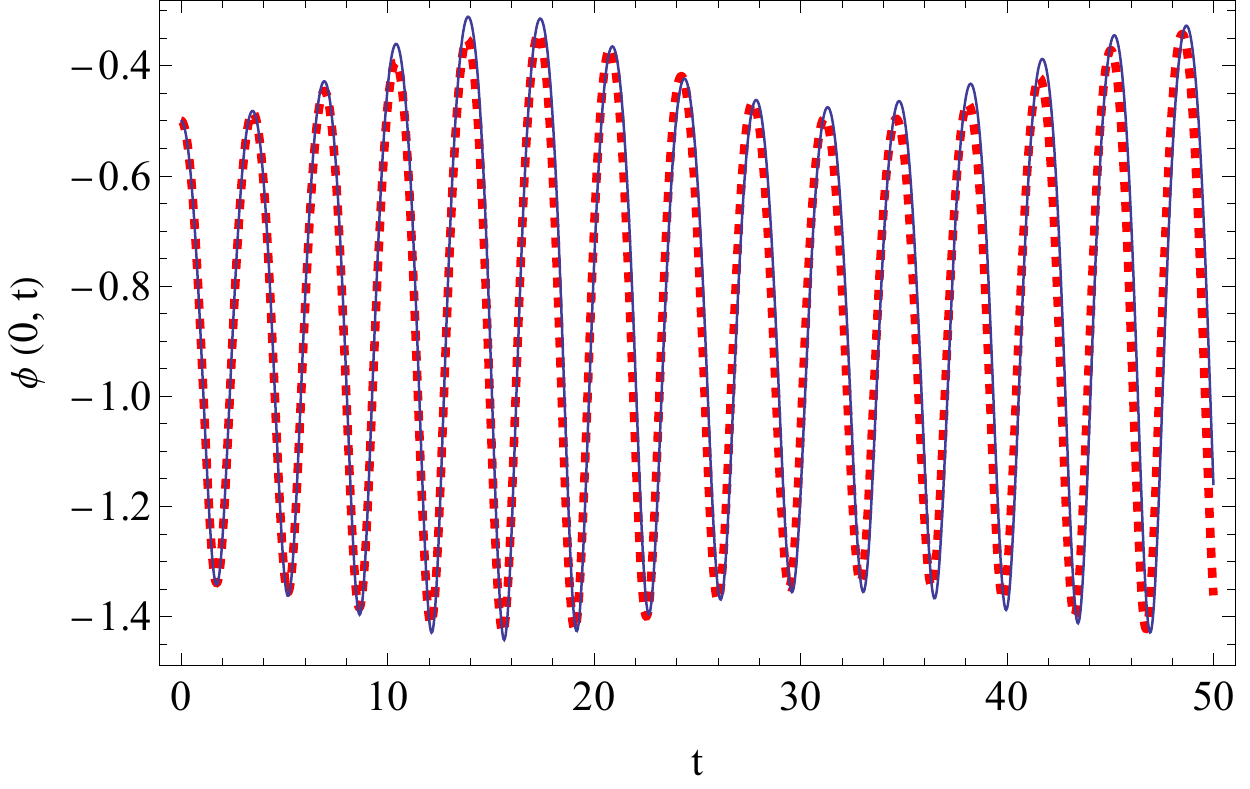} 
\caption{\small Comparison between field theory (solid line) and the effective model with $a_0 = 0.5,\, \delta_0 = 0$, and $R = 2$ (dashed line).}
\label{oscillon_shape}
\end{figure}

Notice that the frequency of the width perturbation $\delta(t)$ is above the threshold. At first glance it may seem that any excitation of $\delta(t)$ should be dissipated very fast through the coupling with radiation. However, as it happens with the oscillon itself, if $\delta(t)$ is excited at sufficiently high amplitudes, the non-linear terms may decrease its frequency below $\omega_t$, becoming a confined mode. Let us now investigate the accuracy of our new effective model with field theory for the initial conditions of Fig. \ref{oscillon_radiation} (d). The corresponding comparison is depicted in Fig. \ref{oscillon_shape}.
Note that the prediction of the effective model (\ref{lag_osc_internal}) and field theory agree with great precision. Therefore, we can confirm that the previous quasi-periodic behaviour is related to the existence of an internal state bounded to the oscillon. Despite the accuracy of the model we can go beyond and include the radiation modes as well, since these degrees of freedom may be significant for certain field configurations. However, as shown in Fig. \ref{oscillon_radiation}, the addition of genuine radiation modes to the effective model does not seem to dissipate efficiently the energy. If one is interested in the dynamics close to the oscillon core and for not very large times, one could add instead modes that resemble the scattering modes, i.e., with spatial frequencies above the mass threshold, but confined to the oscillon. Following this strategy we consider the following ansatz
\be\label{KAK_ansazt_size_2}
\Phi_\text{o, rad}(x;a,\delta)=-1+ a(t) e^{-\left(\frac{x}{R}\right)^2}+\sum_{k=1}^n \delta_k(t) h_k(x/r)\,.
\ee
For simplicity, we have changed the initial oscillon profile to a Gaussian profile. Although (\ref{KAK_ansazt_1}) seems to model better the oscillon  tails, the profile (\ref{KAK_ansazt_size_2}) greatly simplifies the calculations.  We choose the modes as follows
\be
h_k(x/R)=\dfrac{1}{k!}\frac{d^k}{dr^k}e^{-\left(\frac{x}{r}\right)^2}.
\ee

The first one is, in fact, the Derrick mode associated to the new profile. The precise choice of the rest of the modes is not very relevant as long as they have increasing spatial frequencies in the oscillon core. The effective Lagrangian can be written symbolically in a very simple way
\be\label{Lag_os_new}
\mathcal{L}^{o}_{r}=\sum_{k,l = 0}^{n}m_{k,l} \dot{\xi}_k(t)\dot{\xi}_l(t)-\sum_{k,l = 0}^{n}\omega^2_{k,l} \xi_k(t)\xi_l(t)-V(\xi_k(t))\,,
\ee
where $\xi_0(t)=a(t)$ and $\xi_k(t)=\delta_k(t)$ for $k=1,...n$, the matrices $m_{k,l}$ and $\omega^2_{k,l}$ are constant, and $V(\xi_k(t))$ is a potential that couples non-linearly all the modes. Using standard techniques we can diagonalize simultaneously $m_{k,l}$ and $\omega_{k,l}$ and rewrite (\ref{Lag_os_new}) in normal coordinates $\eta_k(t)$ as follows
\be\label{Lag_os_final}
\mathcal{L}^\text{o}_\text{r}=\sum_{k=0}^{n} m_k \dot{\eta}_k^2 (t)-\sum_{k=0}^{n} \omega_k^2 \eta_k^2 (t)-V(\eta_k(t))\,.
\ee
Therefore, in our approach, the oscillon is described by $n+1$ anharmonic oscillators of proper frequencies $\omega_k$ coupled non-linearly by the potential $V(\eta_k(t))$. The system (\ref{Lag_os_final}) is conservative, therefore it cannot dissipate energy. However, the modes $\delta_k(t)$ act effectively as dissipative degrees of freedom, storing energy from the mode amplitude $a(t)$. The energy transfer mechanism works actually very efficiently as we will show.  As a consequence, for not very large times, the model is able to describe how the initial data decays into radiation.
 
In our following experiment we study the decay of an initial configuration of the form 
\be\label{KAK_ansazt_size_3}
\Phi_\text{o}(x;a_0)=-1+ a_0 e^{-\left(\frac{x}{R}\right)^2}\,.
\ee
Below a critical value of $a_0$, the initial configuration given by (\ref{KAK_ansazt_size_3}) decays into radiation. This value can be taken as the minimal $a_0$ such that the proper frequency of a(t) coincides with the mass threshold frequency. By expanding the frequency and imposing that the secular terms vanish, we can compute the frequency correction due to higher order terms. This gives
\be
\omega=\omega_0+\frac{a_0^2 \left(3 \sqrt{2} \omega _0^2-80\right)}{8 \omega _0^3}+\mathcal{O}(a_0^4),\,\, \omega_0=\sqrt{4+1/R^2}\,.
\ee
From this expression we can compute the critical amplitude value for the oscillon formation
\be
a_0^\text{critical}(R)=\frac{\sqrt{\frac{2}{191} \left(20+3 \sqrt{2}\right)}}{R}\,.
\ee

\begin{figure}[H]
\centering
\hspace{-0.5cm}
\subfloat[$a_0 = 0.1$, $R=4$, $r = 4$, $n=7$]{
\includegraphics[scale=0.648]{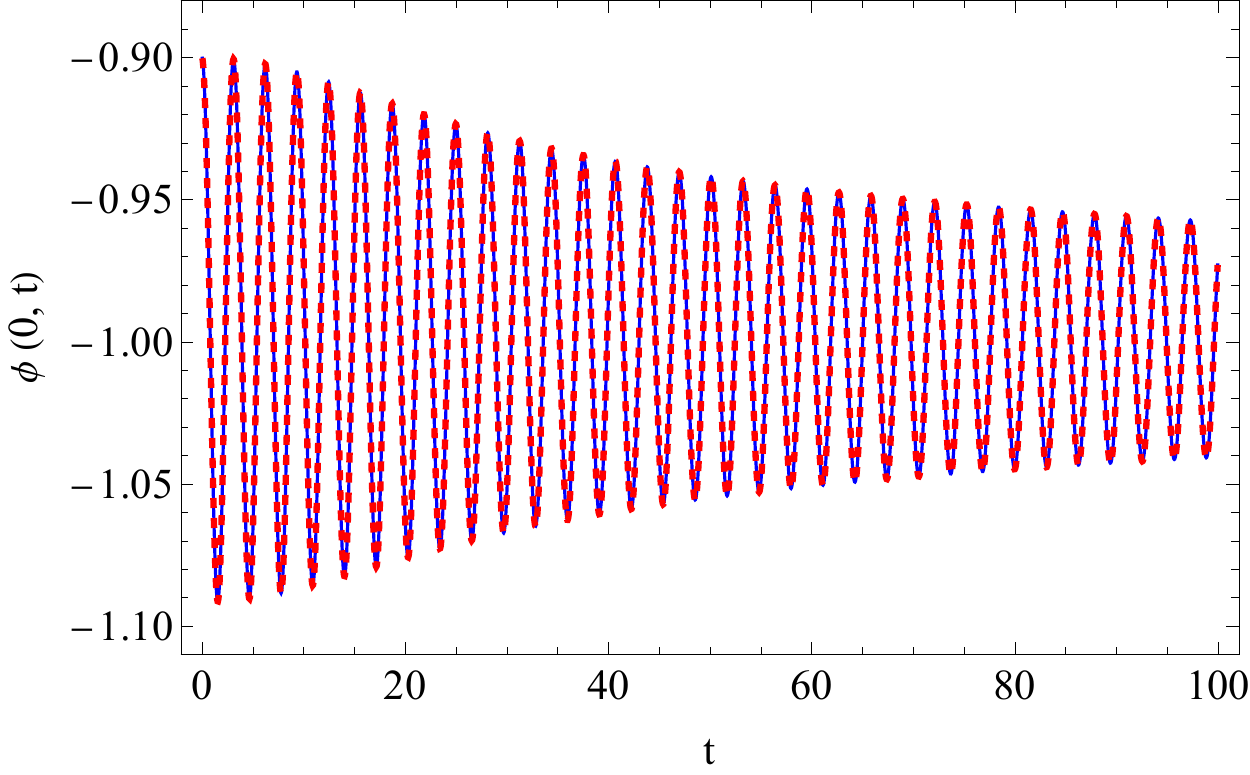}   
}
\subfloat[$a_0 = 0.5$, $R=4$, $r = 4$, $n=7$]{
\includegraphics[scale=0.64]{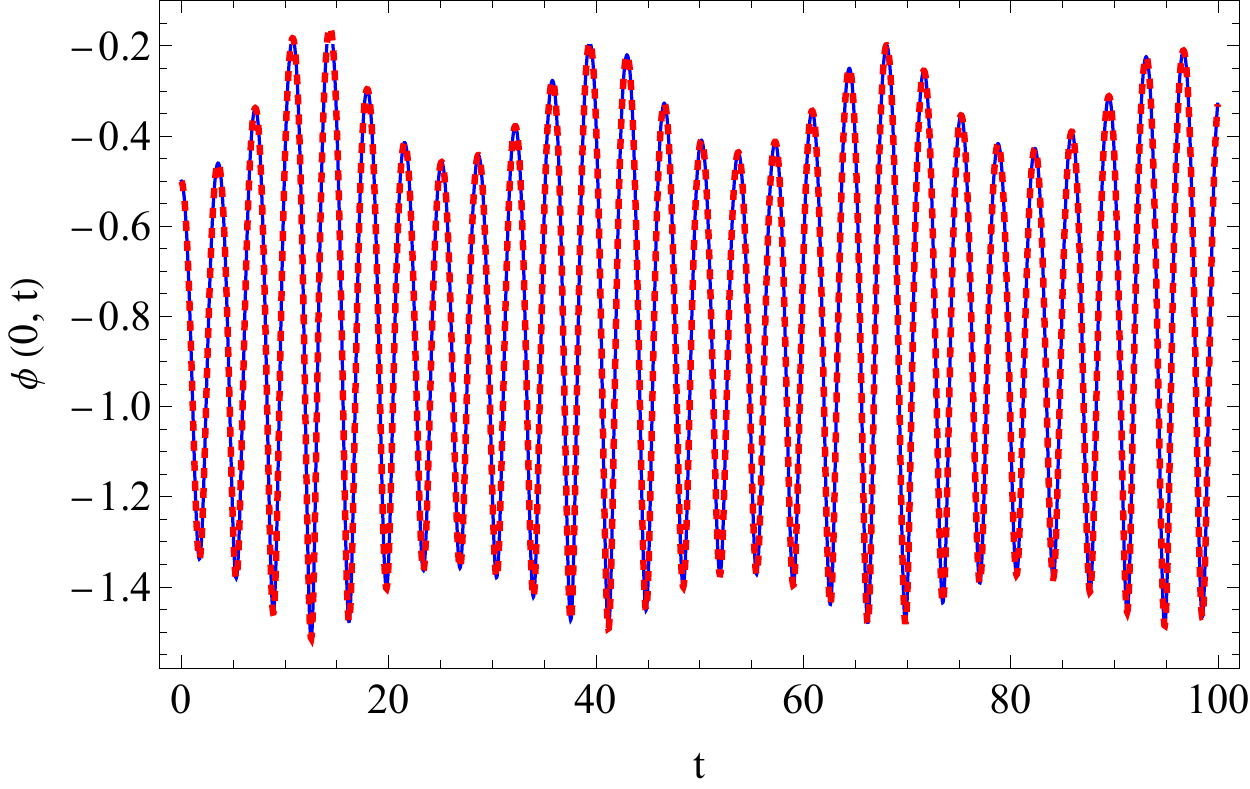}   
}
\caption{\small $\phi(0,t)$ for the initial data  (\ref{KAK_ansazt_size_3}) for different values of the initial amplitude in full numerics (solid line) and in the effective model from (\ref{Lag_os_final}) (dashed line).}
\label{oscillon_internal_full}
\end{figure}

 One situation with $a_0<a_0^\text{critical}(R)$ is illustrated in Fig. \ref{oscillon_internal_full}a. The effective model mimics the decay accurately for $t \lesssim 60$. For $t>60$ the energy stored in the internal modes is transferred back to the amplitude. In Fig. \ref{oscillon_internal_full}b we show a genuine oscillon with an internal mode excited. 

Interestingly, the effective model (\ref{Lag_os_final}) also describes the creation of a $K\bar{K}$ pair from the oscillon profile. In order to understand this phenomenon it is enough to analyse the effective action for $a(t)$. Similarly to (\ref{Lag_eff_osci}), the effective Lagrangian for $a(t)$ with the profile (\ref{KAK_ansazt_size_2}) is given by
\be
\mathcal{L}^o=\sqrt{\frac{\pi }{2}} R \left(\frac{1}{2} \dot{a}(t)^2-\frac{\left(4 R^2+1\right) a(t)^2}{2R^2}-\frac{a(t)^4}{2 \sqrt{2}}+2 \sqrt{\frac{2}{3}} a(t)^3\right).
 \ee
 
\begin{figure}[H]
\begin{center}
\includegraphics[scale=0.75]{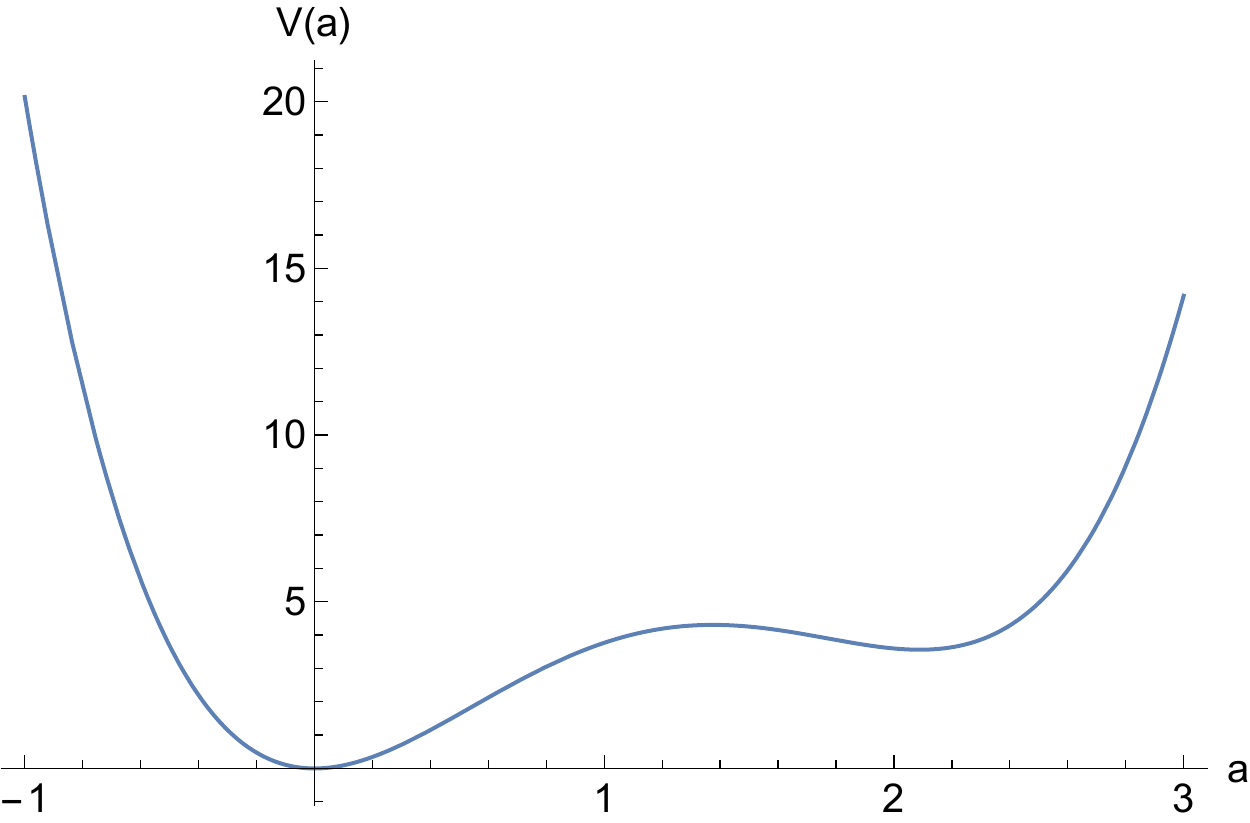}    
\caption{\small Effective potential for $a(t)$ at $R=4$.}
\label{fig:effective_pot}
\end{center}
\end{figure}

The potential for $a(t)$ is depicted in Fig. \ref{fig:effective_pot}. For $R\gtrsim 2.6$ the potential develops a new local minimum around $a\approx 2$. For large enough initial amplitudes, $a(t)$ is able to climb the potential barrier and sit on the upper minimum.
If internal modes are absent and $a(t)$ possesses sufficient energy to overcome the potential barrier, it must descend to the minimum at zero due to the conservation of energy. But when the internal modes are included, they are able to store the energy excess allowing $a(t)$ to oscillate in the upper minimum. This situation is identified with the formation of a $K\bar{K}$ pair which leaves the positive vacuum $\phi=1$ at $x=0$.
 
\begin{figure}[H]
\centering
\subfloat[Full numerics. $R=4$.]{
\includegraphics[scale=0.3775]{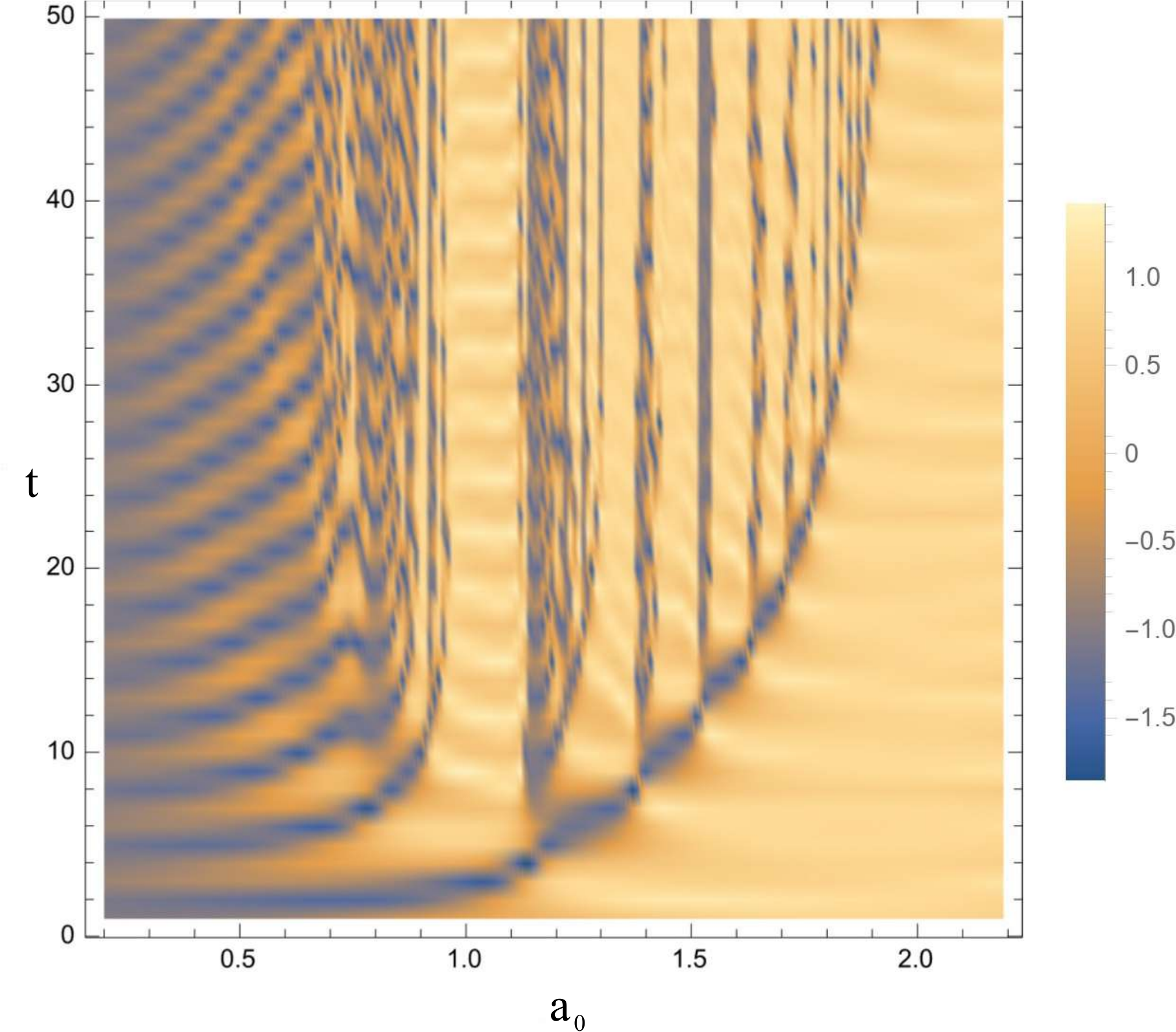}   
}
\subfloat[Effective model. $R=4$, $r=2$ and $n=8$.]{
\includegraphics[scale=0.38]{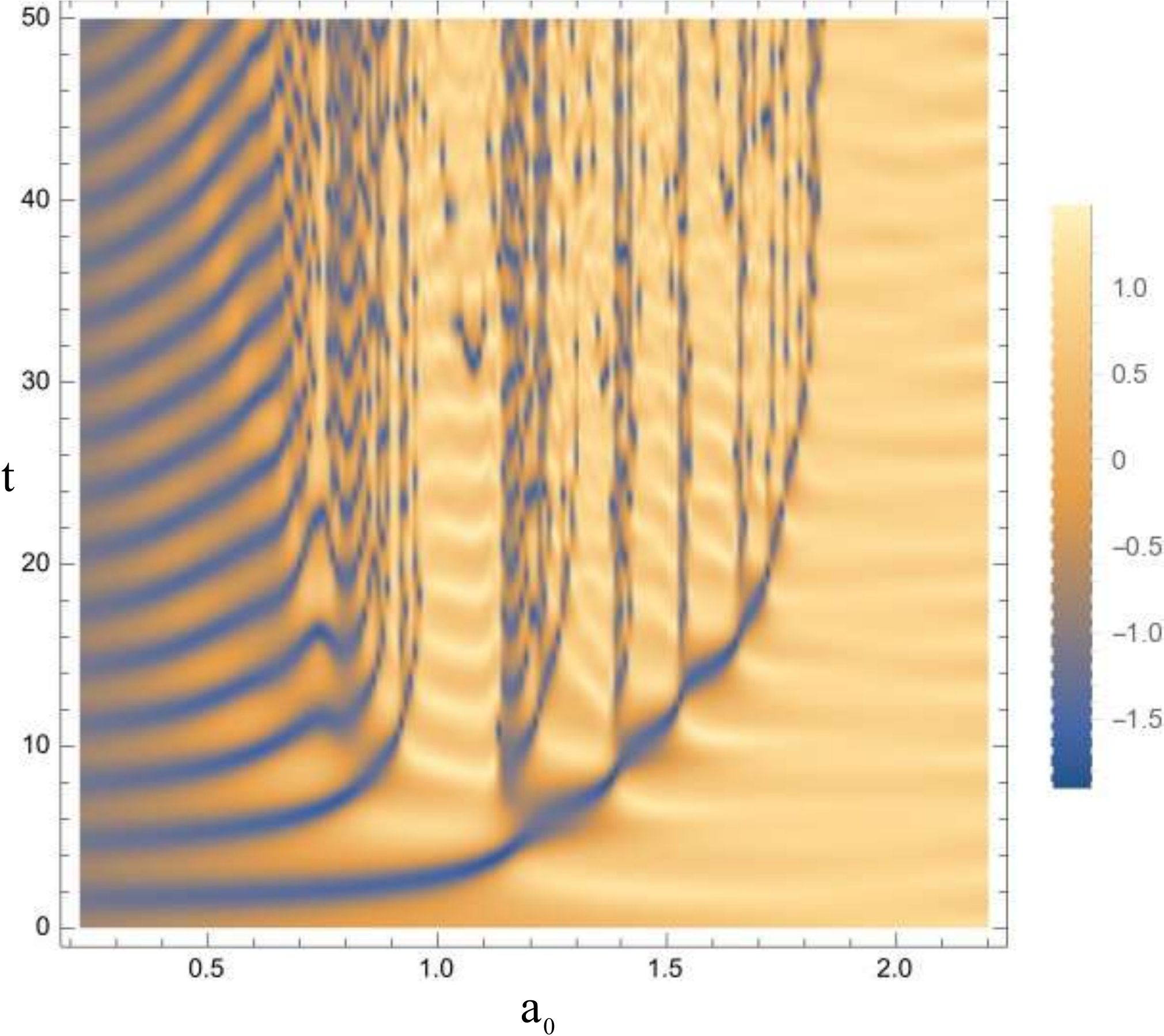}   
}
\caption{\small Comparison between the effective model (\ref{Lag_os_final}) and field theory for the initial data (\ref{KAK_ansazt_size_3}). The color palette indicates the value of the field $\phi$ at the origin, $\phi(0,t)$.}
\label{oscillon_internal_full_1}
\end{figure}

In Fig. \ref{oscillon_internal_full_1} we show the value of the field at the origin $\phi(t,0)$ for different initial amplitudes $a_0$ and $R=4$. The regions where the field takes a value close to $1$ correspond to the $K\bar{K}$ creation. In order to produce a pair the system needs to have an energy $E>2M_k$. But above this value the pair is not always produced, leading to oscillon regions whose energy is dissipated into radiation. This of course resembles the characteristic fractal pattern in the $K\bar{K}$ scattering processes. In the intermediate stages right after the scattering, the field profile looks like a bump above the $-1$ vacuum, resembling the profile of an oscillon. This is the connection between oscillons and $K\bar{K}$ scattering we have mentioned before.  

In order to reproduce the fractal pattern visible at $a_0<2$ one needs to add modes of higher frequencies. This suggests that the scattering modes are essential to explain this structure. Similar results are obtained for values for $  2 \lesssim  R \lesssim 6$. For $R< 2.6$, the effective potential has only a local minima at $a=0$, therefore the field cannot sit on the upper vacuum describing the creation of the $K\bar{K}$ pair. On the other hand, for very large values of $R$, our modes decrease their spatial frequency, and the model as given by the ansatz (\ref{KAK_ansazt_size_2}) does only reproduce qualitatively the pair creation. This can be solved easily by adding higher frequency modes. 

We would like to emphasise that in our approach, the moduli space metric is trivial (i.e. constant), and the relevant dynamics is completely encoded in the specific coupling between modes given by the potential. 

To close this section we are going to illustrate the rich structure given by the internal structure of the oscillons. Apart from the internal mode described by the Derrick mode illustrated in Figs. \ref{oscillon_shape} and \ref{oscillon_internal_full} (b), it is possible to have oscillations with more than one internal mode excited. This gives rise to complicated long-live oscillatory patterns which are very sensitive to the initial conditions.

\begin{figure}[H]
\begin{center}
\includegraphics[scale=0.35]{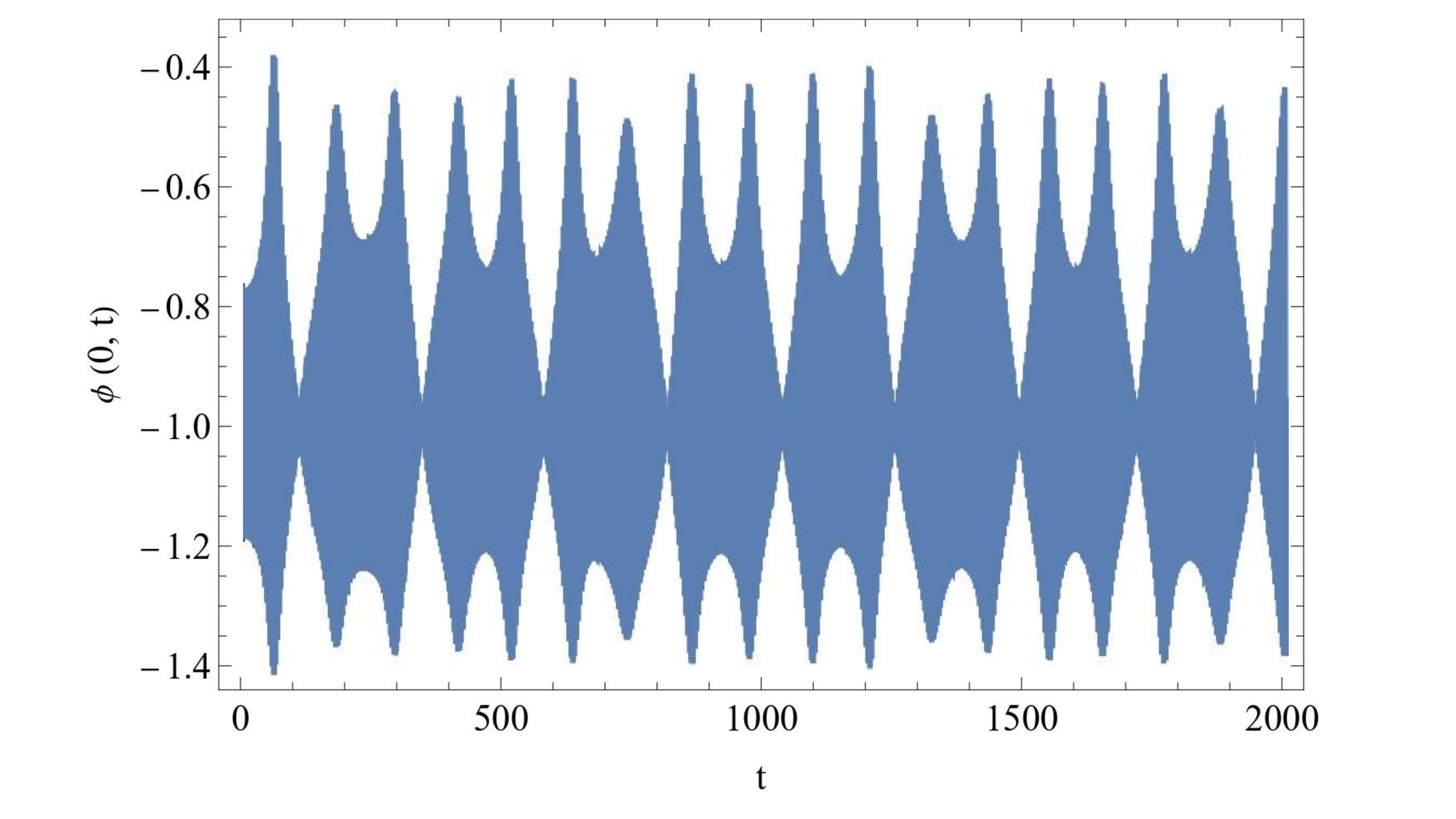}    
\caption{\small $\phi(0,t)$ with initial data given by (\ref{KAK_ansazt_1}) with $R=10$ and $a_0=0.23$.}
\label{fig:two_modes}
\end{center}
\end{figure}

In Fig. \ref{fig:two_modes} we show an interesting pattern. This behaviour is characterized by the excitation of two internal modes. From the point of view of the effective model, the amplitudes of these modes are big enough to decrease their frequencies below the mass threshold. As a consequence, they remain excited for very long times (numerically, at $t\approx 7000$ in our time units, the oscillon ceases to exist). 

The results of this section show the relevance of the radiative modes in the oscillon dynamics. Although for certain initial data the oscillon evolution is well-approximated by the amplitude degree of freedom plus one internal mode, in general, one needs higher frequency modes (which in our approach are confined to the oscillon core) to describe correctly the dynamics. Interestingly, a reasonably simple model (a set of coupled anharmonic oscillators) is able to describe the main features of the oscillon dynamics, including the decay into radiation and the pair $K\bar{K}$ creation.

\section{Summary and Conclusions}
\label{conclusions}
 
In this paper, we have introduced dissipative degrees of freedom in the moduli space approximation of the $\phi^4$ model. We have computed, within this approach, the radiation emitted at infinity by a wobbling kink at the lowest order in the shape mode amplitude. Our results are in complete agreement with the well-know calculations from \cite{Manton2}. 

We begin our investigation by studying in detail the interaction of radiation with the vibrational mode. In terms of the effective model, there are two leading mechanisms that explain the energy transfer between them: a Mathieu instability and a resonance. These mechanisms show that the strongest coupling between radiation and the shape mode occurs for $\omega_q = 2\omega_s$. We contemplate two different experiments: in the first one we analyse a kink with its shape mode initially excited and discuss the main frequency of the radiation emitted. In the second one, we irradiate a kink with linear radiation and study how the shape mode is triggered. Notably, we found an analytic expression for the excitation of the shape mode for frequencies away from the unstable region.
 
We have also studied the role of the translational mode. Despite the fact that the standard CCM approach is non-relativistic, when considered up to second order in $\dot{a}^2(t)$, it is able to reproduce at second order the Lorentz contraction of the kink. The vibrational degree of freedom is not enough to reproduce correctly this relativistic effect, however, the inclusion of scattering modes allows for an exact Lorentz contraction at second order. This suggest that an effective model able to describe dissipative effects should be a nice candidate to describe detailed features of non-linear processes such as kink scattering. Of course, the model looses its usefulness if all radiation modes are included, since basically one recovers field theory. However, a judicious choice of modes describing effectively the scattering modes could shed more light on the understanding of many non-linear processes.

We devoted the last section to the derivation of an effective model for the $\phi^4$ oscillon. Although its natural frequency is above the mass threshold limit, the non-linear terms decrease the frequency below such a limit, avoiding the direct coupling with radiation. The numerical simulations indicate that the oscillon can host a discrete mode responsible for modifications of the width. We implemented this behaviour through the inclusion of the Derrick mode associated to the change of size of the oscillon. This new proposal gives a good agreement with the full numerical simulations. We have added higher frequency modes confined to the oscillon core. They represent scattering modes which may store energy  for certain time, acting effectively as dissipative degrees of freedom.  The effective equations are a system of coupled anharmonic oscillators with a trivial moduli space metric.  Interestingly, once these degrees of freedom are added, this simple effective model is able to describe the KAK creation from initial oscillon data. 
 
Our results suggest that the radiation modes play a crucial role in the study of solitons dynamics. They are of course necessary to explain the decay of non-topological solution such as the oscillon, or long-lived internal modes such as the shape mode. But they also seem to be fundamental to disentangle the complicated patterns in soliton scattering processes. In addition, the results presented here can be easily generalised to other models. The study of the internal structure of oscillons within this approach in different models deserve further research, therefore is left for a future investigation.


\section*{Acknowledgments}

J.Q. thanks J.J. Blanco-Pillado and A. Wereszczynski for useful discussions. This research was supported by Spanish MCIN with funding from European Union NextGenerationEU (PRTRC17.I1) and Consejeria de Educacion from JCyL through QCAYLE project, as well as MCIN project PID2020-113406GB-I00. SNO's research is carried out thanks to a pre-doctoral contract financed by the Junta de Castilla y León through the European Social Fund Plus (ESF+). 

\appendix


\section{Radiation from the shape mode}
\label{AppendixA}

In this appendix we provide some details of the calculation of the radiation emitted by a wobbling kink with its shape mode excited.  We start with (\ref{radiation}) with the scattering amplitudes given by (\ref{eq_cq}) and split it as follows
\begin{equation}\label{eq_rad}
R(x,t) = R_{1}(x) + R_{2}(x,t),
\end{equation}
where
\begin{eqnarray}
R_{1}(x) &=&  i \int_{\mathbb{R}}\, dq\, R_{0}(q)\,\left(4\omega_{s}^{2} - \omega_{q}^{2}\right) \eta_{q}(x)\,, \label{R1}\\
R_{2}(x,t) &=& - i \int_{\mathbb{R}}\, dq\,  R_{0}(q)\, \left(\omega_{q}^{2}\cos(2 \omega_{s} t) + (4\omega_{s}^2 - 2\omega_{q}^{2})\cos(\omega_{q} t)\right) \eta_{q}(x)\, ,   \label{R2}
\end{eqnarray}
and
\be
R_{0}(q) = \dfrac{3 A_{0}^2}{64}\dfrac{ q^2 \left( q^2 - 2 \right)}{ \sqrt{q^2 + 1}\, \omega_{q} (4\omega_{s}^{2} - \omega_{q}^{2})\sinh(\pi q / 2)}\,.
\ee


\subsection{Evaluation of the function $R_{1}(x)$}

\begin{figure}[H]
\begin{center}
\includegraphics[width=0.65\textwidth]{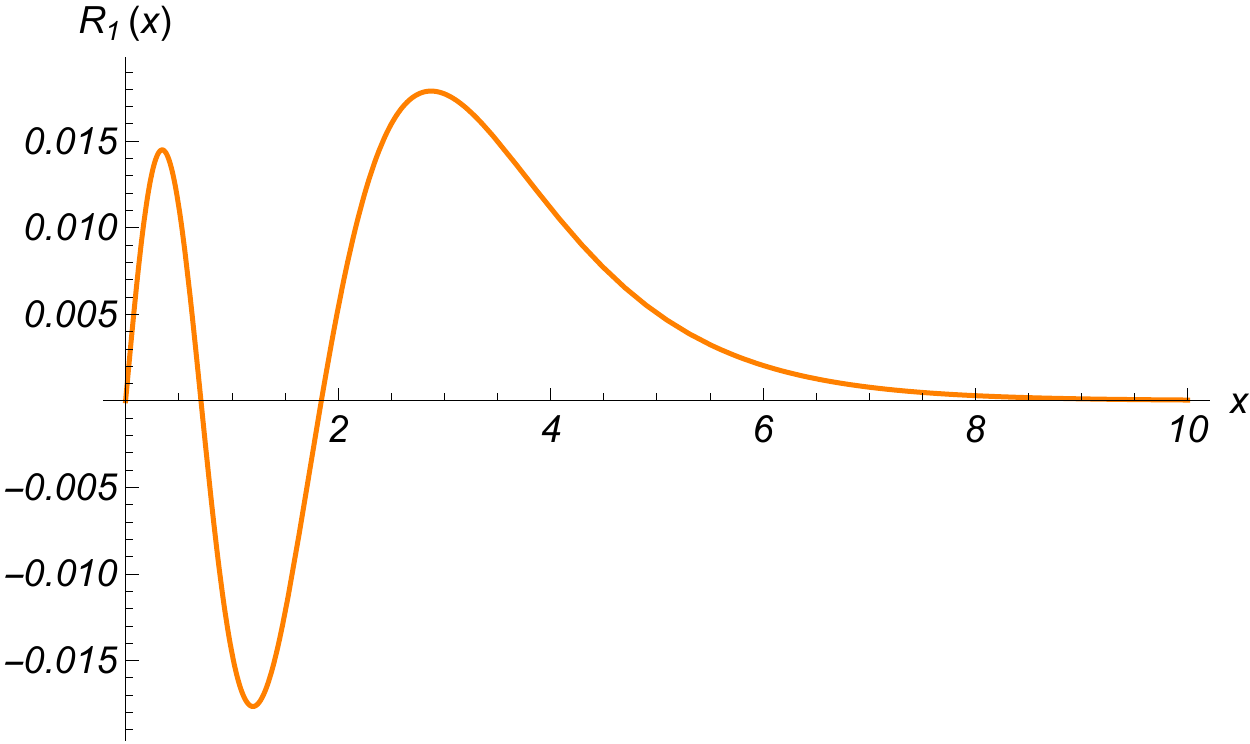}  
\caption{\small Representation of $R_{1}(x)$ given by \eqref{finalR1} for $A_{0} = 1$.}
\label{fig:R1}
\end{center} 
\end{figure}

Let us begin with (\ref{R1}). In the first place, notice that we can divide the integrand into an odd and even contribution. Due to the symmetric interval of integration, only the even contribution is not null, which results in 
\begin{equation}
R_{1}(x) =  - 2 \int_{0}^{\infty}\, dq\, R_{0}(q)\,(4\omega_{s}^{2} - \omega_{q}^{2})\operatorname{Im}(\eta_{q}(x)),
\end{equation}
where 
\begin{equation}
\operatorname{Im}(\eta_{q}(x)) = -( q^2 +1) \sin qx - 3\, q \tanh x \cos qx + 3 \tanh^2 x \sin q x\,.
\end{equation}
Now we will split the integral as
\begin{eqnarray}\label{R1_integrals}
R_{1}(x) &=& \frac{3 A_{0}^{2}}{32}  \bigg( \int_{0}^{\infty}dq\, \dfrac{q^2\,(q^2 - 2)\sin qx}{(q^2 + 4)\sinh(\pi q /2)} + 3\tanh x \int_{0}^{\infty}dq\, \dfrac{q^3\,(q^2 - 2)\cos qx }{(q^2 + 4)(q^2 + 1)\sinh(\pi q / 2)}\nonumber\\ [1ex]
& & - 3 \tanh^2 x \int_{0}^{\infty}dq\, \dfrac{q^2\,(q^2 - 2)\sin qx}{(q^2 + 4)(q^2 + 1)\sinh(\pi q / 2 )} \bigg)\,,
\end{eqnarray} 
and we will solve each term separately by transforming the problem in the calculation of the solution of an inhomogeneous  differential equation. To simplify the notation, let us write
\begin{eqnarray}\label{R1_parts}
R_{1}(x) &=&  \frac{3 A_{0}^{2}}{32}  \bigg( \alpha(x) + 3\tanh x \, \beta(x) - 3 \tanh ^2 x \, \gamma(x) \bigg)\,.
\end{eqnarray} 
The inhomogeneous differential equation satisfied by $\alpha(x)$ is
\begin{equation}
\alpha''(x) - 4 \alpha(x) = - \int_{0}^{\infty} dq\, \dfrac{ q^2 (q^2 - 2) \sin qx}{\sinh(\pi q / 2)} = - 6 \, (3 - \cosh 2x) \tanh x \sech^4 x .
\end{equation}
Given the obvious condition $\alpha(x) = 0$, and to avoid divergences as $\abs{x} \rightarrow \infty$, the solution to this differential equation turns out to be
\begin{equation}\label{alpha:R1}
\alpha(x) = 6 e^{2 x} \log \left( 1+e^{-2 x} \right) - 6 e^{-2 x} \log \left(  1 + e^{2 x}\right) - 2 \tanh  x \left(3 - \sech^2 x  \right)\,.
\end{equation}
Regarding the function $\beta(x)$, it must be solution of the forth order differential equation 
\begin{equation}
\beta^{(4)}(x) - 5\beta''(x) + 4\beta(x) = \int_{0}^{\infty} dq\, \dfrac{ q^3 (q^2 - 2) \cos qx}{\sinh(\pi q / 2)}  = 3 \, (21 - 18 \cosh 2x + \cosh 4x ) \sech^6 x, 
\end{equation}
with conditions $\beta'(0)= 0$ and $\beta^{(3)}(0) = 0$. In order to avoid divergences at $\abs{x} \rightarrow \infty$, the solution must be
\begin{eqnarray}\label{beta:R1}
\beta(x) &=& - 4  e^{2 x} \log  ( 1 + e^{-2 x}  ) - 4 e^{- 2 x} \log  ( 1 + e^{2 x}   ) + 5 - \dfrac{4}{ (1 + e^{2 x}  )^2} + \frac{\pi  e^x}{2} - 2 \tanh x \nonumber\\
& & - 2 \sinh x \arctan e^x .
\end{eqnarray}   
Finally, the $\gamma(x)$ term of $R_{1}(x)$ will be the solution of the differential equation
\begin{equation}
\gamma^{(4)}(x) - 5\gamma''(x) + 4\gamma(x) = \int_{0}^{\infty} dq\, \frac{q^2 (q^2 - 2)}{\sinh( \pi q / 2)}\sin qx = 6 \, (3 - \cosh 2x) \tanh x \sech^4 x, 
\end{equation}
with conditions $\gamma(0) = 0$ and $\gamma''(0) = 0$. The well behaved solution at $\abs{x} \rightarrow \infty $ is
\begin{eqnarray}\label{gamma:R1}
\gamma(x) &=& 2 e^{-2 x} \log  (1+ e^{2 x}) - 2 e^{2 x} \log  (1+  e^{-2 x} ) - e^{-2 x} ( 1 + \tanh x  )  + 1 + \frac{\pi  e^x}{2} \nonumber\\
& & - 2 \cosh x  \arctan e^x.
\end{eqnarray}
Substituting (\ref{alpha:R1}), (\ref{beta:R1}) and (\ref{gamma:R1}) into (\ref{R1_parts}), we get the complete expression of $R_{1}(x)$ 
\begin{eqnarray}\label{finalR1}
R_{1}(x) = \dfrac{3 A_{0}^{2}}{64 \cosh^2 x} \left( 3\pi \sinh x + 16 \tanh x - 24 x \right)\,.
\end{eqnarray}
A plot of this function $R_{1}(x)$ can be seen in Fig.~\ref{fig:R1}.

\subsection{Evaluation of the function $R_{2}(x,t)$}

The evaluation of the contribution of $R_2(x,t)$ seems to be more challenging. Although an analytical calculation seems impossible, some properties of the radiation can be obtained by certain approximations. We will focus now on the asymptotic radiation. As we have seen, the $R_1(x)$ contribution is exponentially suppressed for large $x$, therefore the asymptotic radiation is entirely given by the $R_{2}(x,t)$ contribution. 

Through some elemental algebra is straightforward to verify that the integrand of $R_{2}(x,t)$ can be written as 
\begin{equation}\label{rad_app_1}
(4\omega_{s}^2 - \omega_{q}^2)\bigg(\cos (\omega_q t) - \cos(2 \omega_s t) + \cos(\omega_q t) - 4\omega_{s}^2\dfrac{\cos(\omega_q t) - \cos(2\omega_s t)}{4\omega_{s}^2 - \omega_{q}^2} \bigg)\, ,
\end{equation}
where we have removed temporarily a $-i R_0(q)\hspace{0.01cm}\eta_q(x)$ factor. This expression can be arranged as

\begin{equation}\label{term_parenthesis}
(4\omega_{s}^2 - \omega_{q}^2)\bigg(-2\sin(\dfrac{\omega_q + 2\omega_s}{2}t)\sin(\dfrac{\omega_q - 2\omega_s}{2}t) + \cos(\omega_q t) - \dfrac{8\omega_s^2\sin(\dfrac{\omega_q + 2\omega_s}{2}t)\sin(\dfrac{\omega_q - 2\omega_s}{2}t)}{\omega_{q}^{2} - 4\omega_{s}^{2}} \bigg)\, .
\end{equation}

Let us define the following variables
\begin{equation}
\Omega_q^{+} = \dfrac{\omega_q + 2\omega_s}{2}, \qquad \Omega_q^{-} = \dfrac{\omega_q - 2\omega_s}{2}\, .
\end{equation}
As it is well known, there are some sequence of functions that converge weakly to the Dirac delta. One of them is  
\begin{equation}
f_n(x) = \dfrac{\sin(n x)}{\pi x} \overset{n \rightarrow \infty}{\longrightarrow} \delta(x) .
\end{equation}
In our case, we could identify 
\begin{equation}\label{dirac_Approx}
\dfrac{\sin(\Omega_q^{-} t)}{\pi \Omega_q^{-}} \overset{t \rightarrow \infty}{\longrightarrow} \delta(\Omega_q^{-}) = \sqrt{6}\bigg( \delta(q - 2\sqrt{2}) + \delta(q + 2\sqrt{2}) \bigg)\, ,
\end{equation}
where it was used a property of the Dirac delta. In order to achieve this, let us rewrite (\ref{term_parenthesis}) as
\begin{equation}\label{term_parenthesis:2}
(4\omega_{s}^2 - \omega_{q}^2)\bigg(-2\pi \Omega_q^{-} \sin(\Omega_q^{+}t)\dfrac{\sin(\Omega_q^{-}t)}{\pi \Omega_q^{-}} + \cos(\omega_q t) - 2\pi t\, \omega_s^2\sinc(\Omega_q^{+}t)\dfrac{\sin(\Omega_q^{-} t)}{\pi \Omega_q^{-}}  \bigg)\, .
\end{equation}
Substituting (\ref{dirac_Approx}) into (\ref{term_parenthesis:2}) and integrating, we can realise that the contribution from the first term in (\ref{term_parenthesis:2}) vanishes. This fact is supported by the numerical analysis, which shows that the contribution is suppressed in time. The second term may be computed as follows: let us call such contribution as  
\begin{equation}\label{I_Integral}
\Psi(x,t) = i \int_{\mathbb{R}}\, dq\, R_{0}(q)\, (4\omega_{s}^2 - \omega_{q}^2)\cos(\omega_{q} t)\eta_{q}(x)\, .
\end{equation}
To facilitate the calculation, let's divide $\Psi(x,t)$ in the same way as we did for $R_{1}(x)$
\begin{eqnarray}
\Psi(x,t) &=& \frac{3 i A_{0}^{2}}{64}  \bigg( \int_{\mathbb{R}}dq\, \dfrac{q^2\,(q^2 - 2)\cos(\omega_{q} t)}{(q^2 + 4)\sinh(\pi q /2)}e^{i q x} + 3i\tanh x \int_{\mathbb{R}}dq\, \dfrac{q^3\,(q^2 - 2)\cos(\omega_{q} t)}{(q^2 + 1)(q^2 + 4)\sinh(\pi q / 2)}e^{i q x}\nonumber\\ 
& & - 3 \tanh^2 x \int_{\mathbb{R}}dq\, \dfrac{q^2\,(q^2 - 2)\cos(\omega_{q} t)}{(q^2 + 1)(q^2 + 4)\sinh(\pi q / 2 )}e^{i q x} \bigg)\,.
\end{eqnarray} 
Each of the terms from $\Psi(x,t)$ can be conceived as wave packets where the dispersion relation is non-linear and where the constituent amplitudes are exponentially suppressed in $q$
\begin{eqnarray}\label{def_I}
\Psi(x,t) &=& \frac{3 i A_{0}^{2}}{138}  \bigg( \int_{\mathbb{R}}dq\, \dfrac{q^2\,(q^2 - 2)}{(q^2 + 4)\sinh(\pi q /2)}\left(e^{i (q x + \omega_{q}t)} + e^{i (q x - \omega_{q}t)} \right)\nonumber\\
& & + 3i\tanh x \int_{\mathbb{R}}dq\, \dfrac{q^3\,(q^2 - 2)}{(q^2 + 1)(q^2 + 4)\sinh(\pi q / 2)}\left(e^{i (q x + \omega_{q}t)} + e^{i (q x - \omega_{q}t)} \right)\nonumber\\ 
& & - 3 \tanh^2 x \int_{\mathbb{R}}dq\, \dfrac{q^2\,(q^2 - 2)}{(q^2 + 1)(q^2 + 4)\sinh(\pi q / 2 )}\left(e^{i (q x + \omega_{q}t)} + e^{i (q x - \omega_{q}t)} \right) \bigg)\,.
\end{eqnarray} 
In the literature, the problem of the calculation of wave packets is well known when the amplitude is a Gaussian function, that is, $A(q) = e^{-\alpha^2(q - q_{c})^2}$. For that case, as the amplitude is exponentially suppressed away from the maximum, it is a good approximation to Taylor expand the dispersion relation around $q_{c}$. Then, $\omega(q) \approx q_{c}v_{p} + (q - q_{c})v_{g} + \dfrac{1}{2}\Gamma(q - q_{c})^2 + \dots$ where $v_{p}$ is the phase velocity, $v_{g}$ is the group velocity, and $\Gamma$ is the dispersion parameter. The final result of this approximation is collected by the following expression
\begin{equation}\label{wave_packet}
\int_{\mathbb{R}} dq\, A(q)e^{i(qx \mp \omega(q)t)} \approx \dfrac{\sqrt{2 \pi }}{\sqrt{2\alpha^2 \pm i\Gamma  t}} \exp \left(-\dfrac{1}{2} \left( \dfrac{x \mp v_{g}t}{\sqrt{2\alpha^2 \pm i \Gamma t}} \right)^2 \right)\exp\left(i q_{c}(x \mp v_{p}t)\right).
\end{equation} 
Notice that this integral is also suppressed in time. Finally, using the approximation (\ref{dirac_Approx}) in the last term of (\ref{term_parenthesis:2}), we obtain the only non-vanishing contribution, that looks like 
\begin{eqnarray}
R_{2}(x,t) =&& \dfrac{9 \pi A_0^2}{2 \sqrt{8} \sinh(\sqrt{2} \pi)}\sin(2 \sqrt{3}\,t) \sin (2 \sqrt{2}\,x) + \dfrac{3 \pi A_0^2}{2 \sinh(\sqrt{2}\pi)} \sin(2\sqrt{3}\,t) \cos(2 \sqrt{2}\,x) \tanh x\nonumber\\
&&- \dfrac{3\, \pi A_0^2}{2 \sqrt{8}\sinh(\sqrt{2} \pi)} \sin (2 \sqrt{3}\,t) \sin (2 \sqrt{2}\,x) \tanh^2 x\, . 
\end{eqnarray} 
In the asymptotic spatial regime, the previous expression reduces to
\begin{eqnarray}
R_{2}(x,t) =  \dfrac{3\,\pi A_0^2}{2 \sinh(\sqrt{2}\pi)}\sqrt{\dfrac{3}{8}}\bigg(\cos\big(2\sqrt{3}\,t \mp 2\sqrt{2}\,\abs{x} \mp \delta\big) - \cos\big(2\sqrt{3}\,t \pm 2\sqrt{2}\,\abs{x} \pm \delta\big) \bigg)\,,
\end{eqnarray} 
with
\begin{equation}
\delta = \arctan\sqrt{2}\,.
\end{equation}
Note that the Dirac delta approximation picks a single frequency in the $q-$integral, that is why we have obtained a superposition of two travelling waves of the same frequencies and opposite directions, i.e., a standing wave. Choosing the outgoing wave in the positive direction we get
\be\label{rad_inf2}
R_\infty(x,t)= \dfrac{3\,\pi A_0^2}{2 \sinh(\sqrt{2}\pi)}\sqrt{\frac{3}{8}}\cos\big(2\sqrt{3}\,t - 2\sqrt{2}\,x - \delta\big).
\ee

An alternative derivation of the same result can be achieved as follows. Let us split the integral of the last term in (\ref{rad_app_1}) as follows
\bea\label{I_main}
I(x,t)=&&-4 i \omega_s^2\int_{-\infty}^0 dq \tilde{R}_0(q)\tilde{\eta}_q(x)\frac{\cos\left(\omega_q t\right)-\cos\left(2\omega_s t\right)}{4\omega_s^2-\omega_q^2}e^{i q x}\nonumber\\
&&-4 i \omega_s^2\int_{0}^\infty dq \tilde{R}_0(q)\tilde{\eta}_q(x)\frac{\cos\left(\omega_q t\right)-\cos\left(2\omega_s t\right)}{4\omega_s^2-\omega_q^2}e^{i q x}\nonumber\\
&&\equiv I_1+I_2,
\eea
where
\bea
 \tilde{R}_0(q)&=&\dfrac{3 A_{0}^2}{64}\sqrt{\dfrac{q^2 + 4}{q^2 + 1}}\dfrac{ q^2 \left( q^2 - 2 \right)}{\omega_{q}^2 \sinh(\pi q / 2)}, \\ 
 \tilde{\eta}_q(x)&=&\dfrac{ 3 \tanh^2 x -q^2 - 1 - 3iq\tanh x}{\sqrt{(q^2+1)(q^2+4)}}. 
\eea
Note that values of $q$ close to $0$ as well as large values are suppressed by the function $R_0(q)$. Since the main contribution to the integrals (\ref{I_main}) at large $t$ comes from a neighbourhood of $q_\pm=\pm 2\sqrt{2}$ (notice that the modes corresponding to $q=q_\pm$ grow linearly with time), the frequency $\omega_q$ can be linearly approximated by
\be
\omega_q \approx 2\sqrt{3}\pm \sqrt{\frac{2}{3}}\left(q\mp 2\sqrt{2}\right) + \mathcal{O}\left(q\mp 2\sqrt{2}\right)^2.
\ee
Then, using the linear approximation for $\omega_q$ 
we may approximate $I_2$ by the following expression
\be
I_2(x,t)\approx -4 i  \tilde{R}_0(q_+)\tilde{\eta}_{q_+}(x) \omega_s^2\int_{0}^\infty dq\frac{\cos\left(\widetilde{\omega}_q t\right)-\cos\left(2\omega_s t\right)}{4\omega_s^2-\omega_q^2}e^{i q x}\,,
\ee
where $\widetilde{\omega_q} = 2\sqrt{3} +  \sqrt{\frac{2}{3}}\left(q - 2\sqrt{2}\right)$. We are assuming implicitly that in a neighbourhood of $q = 2\sqrt{2}$, both $\tilde{R}_0(q)$ and $\tilde{\eta}_{q}(x)$ are approximately constant in $q$. Similarly, a straightforward manipulation leads to the following expression for $I_1$
\be
I_1(x,t)\approx 4 i  \tilde{R}_0(q_+)\tilde{\eta}^\ast_{q_+}(x) \omega_s^2\int_{0}^\infty dq\frac{\cos\left(\widetilde{\omega}_q t\right)-\cos\left(2\omega_s t\right)}{4\omega_s^2-\omega_q^2}e^{-i q x}\,.
\ee 

We have finally
\be\label{rad_I_1}
I(x,t)=8 \omega_s^2\tilde{R}_0(q_+)\left(\text{Re}\left(\tilde{\eta}_{q_+}(x)\right)\mathcal{F}_s(t)+\text{Im}\left(\tilde{\eta}_{q_+}(x)\mathcal{F}_c(t)\right)\right)\,,
\ee
where
\bea\label{four_1}
\mathcal{F}_s(t)&=& \int_0^\infty dq \frac{\cos\left(\widetilde{\omega}_q t\right)-\cos\left(2\omega_s t\right)}{4\omega_s^2-\omega_q^2}\sin\left(q x\right),\\\label{four_2}
\mathcal{F}_c(t)&=& \int_0^\infty dq \frac{\cos\left(\widetilde{\omega}_q t\right)-\cos\left(2\omega_s t\right)}{4\omega_s^2-\omega_q^2}\cos\left(q x\right)\,.
\eea
The integrals (\ref{four_1}) and (\ref{four_2}) can be computed analytically for all $x$, but the expressions are not particularly illuminating. For $x>>0$ and large $t$ we have
\bea\label{fsin}
\mathcal{F}_s(t)&=& - \dfrac{\pi  \cos \left(2 \sqrt{3}\,t-2 \sqrt{2}\, x\right)}{4 \sqrt{2}}\,,\\\label{fcos}
\mathcal{F}_c(t)&=& - \dfrac{\pi  \sin \left(2 \sqrt{3}\,t-2 \sqrt{2}\, x\right)}{4 \sqrt{2}}\,.
\eea

Substituting (\ref{fsin}) and (\ref{fcos}) into (\ref{rad_I_1})  we obtain (for $x>\!\!>0$)
\be
I(x,t)= \dfrac{3\,\pi A_0^2}{2 \sinh(\sqrt{2}\pi)}\sqrt{\dfrac{3}{8}}\cos\big(2\sqrt{3}\,t - 2\sqrt{2}\,x - \arctanh \sqrt{2} \big)\,.
\ee


\section{Details of the numerical simulations}
\label{AppendixB}

In this appendix we discuss the numerical coding scheme that we have used to solve numerically the equation of motion
\begin{equation}\label{eq_field}
\ddot{\phi} - \phi'' + 2\phi(\phi^2 - 1) = 0\,,
\end{equation}
where dots and primes denote derivatives with respect to time and space respectively. In order to solve (\ref{eq_field}) we have discretized the second order spatial derivative as
\begin{equation}
\phi''(x,t) = \dfrac{\phi(x + \Delta x,t) - 2\phi(x,t) + \phi(x - \Delta x, t)}{(\Delta x)^2}\, ,
\end{equation}
and performed the temporal evolution through the leapfrog method. In addition, we have employed absorbing boundary conditions to avoid that the radiation is scattered-back and interferes with the system. These conditions read as
\begin{eqnarray}
(\partial_t + \partial_x \phi)\Big|_{x = \frac{L}{2},t} &=& 0\,,\\
(\partial_t - \partial_x \phi)\Big|_{x = - \frac{L}{2},t} &=& 0\,.
\end{eqnarray}
The simulations have been run in a lattice of length $L = 30$ with $\Delta x = 0.0375$, so the number of spatial points is $N_{space} = 1600$. Finally, to guarantee the stability of the method we have chosen $\Delta t = \dfrac{L}{10(N_{space} + 2)} $\,, which satisfies the Courant condition with $\Delta t/\Delta x<0.5$.



\begin{thebibliography}{99}

\bibitem{Rajaraman} R. Rajaraman, Solitons and Instantons, Elsevier Science,
Amsterdam (1982).

\bibitem{Manton} N. Manton and P. Sutcliffe, Topological Solitons,
Cambridge University Press, Cambridge U.K. (2004).

\bibitem{Shellard} A. Vilenkin and E. P. S. Shellard, Cosmic Strings and Other Topological Defects, pp. 578, Cambridge U.K.: Cambridge University Press (2000).

\bibitem{Shnir} Y. M. Shnir, Topological and Non-Topological Solitons
in Scalar Field Theories, Cambridge University Press,
Cambridge U.K. (2018).

\bibitem{Sugiyama} T. Sugiyama, ``Kink-antikink collisions in the two dimensional $\phi^4$ model'', Prog. Theor. Phys. \textbf{61} (1979)  1550.

\bibitem{Campbell} D. K. Campbell, J. F. Schonfeld, and C. A. Wingate, ``Resonance structure in kink-antikink interactions in $\phi^4$ theory'', Physica \textbf{D9} (1983) 1.

\bibitem{Relativistic} C. Adam, N.S. Manton, K. Oles, T. Romanczukiewicz and A. Wereszczynski, ``Relativistic moduli space for kink collisions",  Phys. Rev. D \textbf{105} (2022)  065012.

\bibitem{Relativistic1} C. Adam, D. Ciurla, K. Oles, T. Romanczukiewicz and A. Wereszczynski, ``Relativistic Moduli Space and critical velocity in kink collisions",  arXiv: 2304.14076 [hep-th].

\bibitem{Segur}H. Segur and M.D. Kruskal, ``Nonexistence of Small Amplitude Breather Solutions in $\phi^4$ theory", Phys. Rev. Lett. \textbf{58} (1987) 1158.

\bibitem{Hindmarsh2} P. Salmi and M. Hindmarsh, ``Radiation and Relaxation of Oscillons", Phys. Rev. \textbf{D85} (2012) 085033.

\bibitem{Hindmarsh3} M. Hindmarsh and P. Salmi, ``Numerical investigations of oscillons in 2 dimensions," Phys. Rev. \textbf{D74} (2006) 105005.

\bibitem{Fodor1} G. Fodor, P. Forgács, Z. Horváth and A. Lukács, ``Small amplitude quasi-breathers and oscillons", Phys. Rev. \textbf{D78} (2008) 025003.

\bibitem{Fodor3} G. Fodor, P. Forgács, Z. Horváth and M. Mezei, ``Radiation of scalar oscillons in 2 and 3 dimensions", Phys. Lett. \textbf{B674} (2009) 319.

\bibitem{Manton3}  N.S. Manton and T. Romańczukiewicz, ``Simplest oscillon and its sphaleron", Phys. Rev. \textbf{D107} (2023)  085012.

\bibitem{Olle}J. Ollé, O. Pujolàs, and F. Rompineve, “Oscillons and Dark Matter”, JCAP \textbf{02} (2020) 006.

\bibitem{Kawasaki}M. Kawasaki, W. Nakano, and E. Sonomoto, “Oscillon of Ultra-Light Axion-like
Particle”, JCAP \textbf{01} (2020) 047.

\bibitem{Arvanitaki}A. Arvanitaki, S. Dimopoulos, M. Galanis, L. Lehner, J. O. Thompson, and K.
Van Tilburg, “Large-misalignment mechanism for the formation of compact axion structures: Signatures
from the QCD axion to fuzzy dark matter”, Phys. Rev. \textbf{D101} (2020) 083014.

\bibitem{Hindmarsh5}M. Hindmarsh and P. Salmi, “Oscillons and domain walls”, Phys. Rev. \textbf{D77} (2008) 105025.

\bibitem{Gorghetto}M. Gorghetto, E. Hardy, and G. Villadoro, “More axions from strings”, SciPost Phys.
\textbf{10} (2021) 050.

\bibitem{Blanco}J. J. Blanco-Pillado, D. Jiménez-Aguilar, and J. Urrestilla, “Exciting the domain wall soliton”,
JCAP \textbf{01} (2021) 027.

\bibitem{dissel} F. van Dissel, O. Pujolàs and E. I. Sfakianakis, arXiv: 2303.1602 [hep-th].

\bibitem{Manton1} N. S. Manton, K. Oleś, T. Romańczukiewicz and A. Wereszczyński, ``Kink moduli spaces: Collective coordinates reconsidered'', Phys.Rev. \textbf{D103} (2021) 025024.

\bibitem{Manton2} N. S. Manton and H. Merabet, ``$\phi^4$ kinks: Gradient flow and dynamics'', Nonlinearity \textbf{10} (1997) 3.

\bibitem{Tom2}T. Romanczukiewicz, ``Creation of kink and antikink pairs forced by radiation", J. Phys. A: Math. Gen. \textbf{39} (2006) 3479.

\bibitem{Roman}P. Forgács, A. Lukács and T. Romańczukiewicz, ``Negative radiation pressure exerted on kinks", Phys. Rev. \textbf{D77} (2008) 125012.

\bibitem{Rice}M. J. Rice, ``Physical dynamics of solitons", Phys. Rev. \textbf{B28} (1983) 3587.

\bibitem{Fodor2}  G. Fodor, P.  Forgács, Z. Horváth and A. Lukács, Phys. Rev. \textbf{D79} (2009) 065002, ``Computation of the radiation amplitude of oscillons",  Phys. Rev. \textbf{D85} (2012) 043501.

\bibitem{Hindmarsh} P. Salmi and  M. Hindmarsh, ``Radiation and Relaxation of Oscillons",  Phys. Rev. \textbf{D85} (2012) 043501.

\bibitem{Galvez}J. T. Gálvez Ghersi and J. N. Braden, ``Dimensional deformation of sine-Gordon breathers into oscillons", arXiv: 2303.04750 [hep-th].

\bibitem{q1} J. J. Blanco-Pillado, D. Jiménez-Aguilar, J. M. Queiruga and J. Urrestilla, ``Parametric resonances in axionic cosmic strings", JCAP {\bf 04} (2023) 043. 

\bibitem{q2} J. J. Blanco-Pillado, D. Jiménez-Aguilar, J. M. Queiruga and J. Urrestilla, ``The dynamics of Domain Wall Strings", arXiv: 2209.12945 [hep-th].

\end{thebibliography}
\end{document}